\documentclass[fleqn,usenatbib]{mnras}

\usepackage{newtxtext,newtxmath}

\usepackage[T1]{fontenc}
\usepackage{ae,aecompl}
\usepackage{siunitx}
\usepackage{xspace}

\usepackage{graphicx}	
\usepackage{amsmath}	
\usepackage{cancel}
\usepackage[dvipsnames]{xcolor}

\newcommand{\lsim}{\lesssim}

\newcommand{\ALLWISE}{\uppercase{allwise}\xspace}
\newcommand{\ALLMASK}{\uppercase{allmask}\xspace}

\newcommand{\BAO}{\uppercase{bao}\xspace}
\newcommand{\BITMASK}{\uppercase{bitmask}\xspace}
\newcommand{\BGSB}{\uppercase{bgs bright}\xspace}
\newcommand{\BGSF}{\uppercase{bgs faint}\xspace}
\newcommand{\BGS}{\uppercase{bgs}\xspace}
\newcommand{\BASS}{\uppercase{bass}\xspace}

\newcommand{\BTS}{\uppercase{bts}\xspace}
\newcommand{\BS}{\uppercase{bs}\xspace}

\newcommand{\CC}{\uppercase{cc}s\xspace}
\newcommand{\CCDs}{\uppercase{ccd}s\xspace}
\newcommand{\CORRFLUX}{{\uppercase{flux/mw\_transmission}}\xspace}
\newcommand{\CP}{\uppercase{cp}\xspace}
\newcommand{\DECam}{\uppercase{dec}am\xspace}

\newcommand{\DECaLS}{\uppercase{dec}a\uppercase{ls}\xspace}
\newcommand{\DESI}{\uppercase{desi}\xspace}
\newcommand{\DES}{\uppercase{des}\xspace}
\newcommand{\DESITARGET}{\uppercase{desitarget}\xspace}
\newcommand{\DReight}{\uppercase{dr8}\xspace}
\newcommand{\DRtwo}{\uppercase{dr2}\xspace}
\newcommand{\DTS}{\uppercase{dts}\xspace}

\newcommand{\ELGs}{\uppercase{elg}s\xspace}

\newcommand{\EBV}{\uppercase{ebv}\xspace}
\newcommand{\FIBERFLUX}{{\uppercase{fiberflux}}\xspace}
\newcommand{\FLUX}{{\uppercase{flux}}\xspace}
\newcommand{\FLUXIVAR}{{\uppercase{ flux\_ivar}}\xspace}
\newcommand{\FMC}{{\uppercase{fmc}}\xspace}
\newcommand{\FRACMASKED}{{\uppercase{fracmasked}}\xspace}
\newcommand{\FRACFLUX}{{\uppercase{fracflux}}\xspace}
\newcommand{\FRACIN}{{\uppercase{fracin}}\xspace}
\newcommand{\GAMA}{\uppercase{gama}\xspace}
\newcommand{\GAIA}{{\it Gaia}\xspace}

\newcommand{\GC}{\uppercase{gc}\xspace}

\newcommand{\LBNL}{\uppercase{lbnl}\xspace}

\newcommand{\LRGs}{\uppercase{lrg}s\xspace}
\newcommand{\LS}{\uppercase{ls}\xspace}
\newcommand{\LG}{\uppercase{lg}\xspace}
\newcommand{\SGA}{\uppercase{sga}\xspace}

\newcommand{\MGS}{\uppercase{mgs}\xspace}
\newcommand{\MWTRANSMISSION}{{\uppercase{mw\_transmission}}\xspace}
\newcommand{\MXXL}{\uppercase{mxxl}\xspace}
\newcommand{\MzLS}{\uppercase{m}z\uppercase{ls}\xspace}
\newcommand{\MASKBITS}{\uppercase{maskbits}\xspace}

\newcommand{\NOAO}{\uppercase{NSF}’s \uppercase{OIR} \uppercase{L}ab\xspace}

\newcommand{\NOBS}{\uppercase{nobs}\xspace}
\newcommand{\NGC}{\uppercase{ngc}\xspace}
\newcommand{\NOBSG}{\uppercase{nobs\_g}\xspace}
\newcommand{\NOBSR}{\uppercase{nobs\_r}\xspace}
\newcommand{\NOBSZ}{\uppercase{nobs\_z}\xspace}
\newcommand{\NQ}{\uppercase{nq}\xspace}
\newcommand{\PSF}{\uppercase{psf}\xspace}
\newcommand{\PSFs}{\uppercase{psf}s\xspace}

\newcommand{\QSOs}{\uppercase{qso}s\xspace}
\newcommand{\QCs}{\uppercase{qc}s\xspace}
\newcommand{\RSD}{\uppercase{rsd}\xspace}
\newcommand{\RPETRO}{{\uppercase{ r\_petro}}\xspace}
\newcommand{\SDSS}{\uppercase{sdss}\xspace}
\newcommand{\SGC}{\uppercase{sgc}\xspace}
\newcommand{\SG}{\uppercase{sg}\xspace}
\newcommand{\TRACTOR}{\uppercase{tractor}\xspace}
\newcommand{\TRACTORs}{\uppercase{tractor}s\xspace}
\newcommand{\TYCHO}{Tycho-2\xspace}
\newcommand{\Tycho}{Tycho-2\xspace}
\newcommand{\WCS}{\uppercase{wcs}\xspace}
\newcommand{\Wone}{\uppercase{w1}\xspace}
\newcommand{\Wtwo}{\uppercase{w2}\xspace}

\newcommand{\LEGACYPIPE}{\textsc{legacypipe}\xspace}
\newcommand{\NANOMAGIES}{\uppercase{nanomaggies}\xspace}
\newcommand{\NANOMAGIE}{\uppercase{nanomaggie}\xspace}

\newcommand{\REX}{\uppercase{rex}\xspace}
\newcommand{\DEV}{\uppercase{dev}\xspace}
\newcommand{\EXP}{\uppercase{exp}\xspace}
\newcommand{\COMP}{\uppercase{comp}\xspace}

\newcommand{\SER}{\uppercase{ser}\xspace}
\newcommand{\LCDM}{\uppercase{lcdm}\xspace}

\title[Target selection for the DESI BGS]{Characterising the target selection pipeline for the Dark Energy Spectroscopic Instrument Bright Galaxy Survey}

\author[Omar A. Ruiz-Macias et al.]{Omar Ruiz-Macias,$^{1,2}$\thanks{E-mail: omar.a.ruiz-macias@durham.ac.uk} Pauline Zarrouk,$^{1}$ Shaun Cole,$^{1}$ Carlton M. Baugh,$^{1,2}$
\newauthor{Peder Norberg,$^{1,3}$ John Lucey,$^3$ Arjun Dey,$^{4}$ Daniel J. Eisenstein,$^{5}$ Peter Doel,$^{6}$} 
\newauthor{Enrique Gazta\~naga,$^{7}$ ChangHoon Hahn,$^{8,9}$ Robert Kehoe,$^{10}$ Ellie Kitanidis,$^{11}$}
\newauthor{Martin Landriau,$^{8}$ Dustin Lang,$^{12,13}$ John Moustakas,$^{14}$ Adam D.\ Myers,$^{15}$ }
\newauthor{Francisco Prada,$^{16}$ Michael Schubnell,$^{17}$ David H. Weinberg,$^{18}$ M. J. Wilson,$^{8,9}$}
\\
\scriptsize $^{1}$ Institute for Computational Cosmology, Department of Physics, Durham University, South Road, Durham DH1 3LE, UK\\
\scriptsize $^{2}$ Institute for Data Science, Durham University, South Road, Durham DH1 3LE, UK\\
\scriptsize $^{3}$ Centre for Extragalactic Astronomy, Department of Physics, Durham University, South Road, Durham DH1 3LE, UK\\
\scriptsize $^{4}$ NSF’s National Optical-Infrared Astronomy Research Laboratory, 950 N. Cherry Ave., Tucson, AZ 85719, USA\\
\scriptsize $^{5}$ Harvard-Smithsonian Center for Astrophysics, 60 Garden Street, Cambridge, MA 02138, USA\\
\scriptsize $^{6}$ Department of Physics \& Astronomy, University College London, Gower Street, London, WC1E 6BT, UK\\
\scriptsize $^{7}$ Institute of Space Sciences (ICE, CSIC), Campus UAB, Carrer de Can Magrans, s\/n, 08193 Bellaterra (Barcelona), Spain\\
\scriptsize $^{8}$ Lawrence Berkeley National Laboratory, One Cyclotron Road, Berkeley, CA 94720, USA\\
\scriptsize $^{9}$ Berkeley Center for Cosmological Physics, UC Berkeley, CA 94720, USA\\
\scriptsize $^{10}$ Department of Physics, Southern Methodist University, 3215 Daniel Avenue, Dallas, TX 75205, USA\\
\scriptsize $^{11}$ Department of Physics, University of California, Berkeley, 366 LeConte Hall, Berkeley, CA 94720, USA\\
\scriptsize $^{12}$ Perimeter Institute for Theoretical Physics, 31 Caroline Street N, Waterloo, Ontario, N2L 2Y5, Canada\\
\scriptsize $^{13}$ Department of Physics and Astronomy, University of Waterloo, Waterloo, ON N2L 3G1, Canada\\
\scriptsize $^{14}$ Department of Physics and Astronomy, Siena College, 515 Loudon Road, Loudonville, NY 12211\\
\scriptsize $^{15}$ University of Wyoming, 1000 E. University Ave., Laramie, WY 82071, USA\\
\scriptsize $^{16}$ Instituto de Astrofisica de Andaluc\'ia, Glorieta de la Astronom\'ia, s/n, E-18008 Granada, Spain\\
\scriptsize $^{17}$ Department of Physics, University of Michigan, Ann Arbor, MI 48109, USA\\
\scriptsize $^{18}$ Department of Astronomy and the Center for Cosmology and Astroparticle Physics, The Ohio State University, 140 West 18th Avenue, Columbus OH 43210, USA
}

\date{Accepted 2021 January 26. Received 2021 January 20; in original form 2020 July 30}

\pubyear{2021}

\begin{document}
\label{firstpage}
\pagerange{\pageref{firstpage}--\pageref{lastpage}}
\maketitle

\begin{abstract}
We present the steps taken to produce a reliable and complete input galaxy catalogue for the Dark Energy Spectroscopic Instrument (\DESI) Bright Galaxy Sample (\BGS) using the photometric Legacy Survey \DReight \DECam. We analyze some of the main issues faced in the selection of targets for the \DESI \BGS, such as star-galaxy separation, contamination by fragmented stars and bright galaxies. Our pipeline utilizes a new way to select \BGS galaxies using \GAIA photometry and we implement geometrical and photometric masks that reduce the number of spurious objects. The resulting catalogue is cross-matched with the Galaxy and Mass Assembly (\GAMA) survey to assess the completeness of the galaxy catalogue and the performance of the target selection. We also validate the clustering of the sources in our \BGS catalogue by comparing with mock catalogues and \SDSS data. Finally, the robustness of the \BGS selection criteria are assessed by quantifying the dependence of the target galaxy density on imaging and other properties. The largest systematic correlation we find is a $7$ per cent supression of the target density in regions of high stellar density. 
\end{abstract}




\section{Introduction}\label{sec:intro} 

The Dark Energy Spectroscopic Instrument\footnote{\url{http://desi.lbl.gov/}} (\DESI) \citep{DESI2016:surveys} is a multi-fibre spectrograph that will be used to carry out a number of wide-field surveys of galaxies and quasars to map the large-scale structure of the Universe. These surveys will probe the form of dark energy by allowing high precision measurements of the baryon acoustic oscillation (\BAO) scale and the growth rate of structure using redshift-space distortions (\RSD). The characterisation and definition of the target list for each \DESI survey is a critical step for efficient survey execution and to allow reliable measurements of galaxy clustering. Here we describe this process for the \DESI bright galaxy survey (hereafter \BGS), a flux limited sample of around 10 million galaxies, using photometry from a new imaging survey, the Legacy Surveys\footnote{\url{http://legacysurvey.org/}} (\LS).

\DESI is a robotically-actuated, fibre-fed spectrograph that is capable of collecting $5\,000$ spectra simultaneously.

The spectra cover the wavelength range $360$ to $980$~nm,
with a spectral resolution of $R = \lambda/\Delta\lambda$ between $2\,000$ and $5\,500$, depending on the wavelength. \DESI will be used to conduct a five-year survey starting in 2020, with the aim of measuring redshifts over a solid angle of $14\,000$ $\textrm{deg}^2$. More than 30 million spectroscopic targets will be selected for four different tracer samples drawn from the imaging data. These are (i)  luminous red galaxies (\LRGs) in the redshift range $z = 0.3$ to $z = 1$, (ii) emission line galaxies (\ELGs) to $z = 1.7$, (iii) quasars to higher redshifts ($2.1 < z < 3.5$), and (iv) a magnitude-limited \BGS out to $z \approx 0.6$ with a median redshift of $z \approx 0.2$ which is the focus of this paper.  

\DESI observations are divided into two main programmes: the Bright Time Survey (\BTS) and the Dark Time Survey (\DTS). The \BGS will be part of the \BTS and is conducted when the Moon is above the horizon and the sky is too bright to allow efficient observation of fainter targets. The \BTS excludes the few nights closest to full Moon and \BGS always targets fields that are at least $ 40 - 50$ deg away from the Moon. \BGS alone will be ten times larger than the \SDSS-I and \SDSS-II main galaxy samples (\MGS) of 1 million bright galaxies that were observed over the time period $1999-2008$ \citep{Abazajian:2003jy}.

The target sample for the \BGS is intended to be a galaxy sample that is flux-limited in the $r$-band. The magnitude limit is determined by the total amount of  bright observing time and the exposure times required to achieve the desired redshift efficiency. This target selection is, in essence, a deeper version of the target selection for the \SDSS \MGS \cite[]{2002AJ....124.1810S}.

To make predictions for \BGS target sample we make use of the mock galaxy catalogue created from the Millennium-XXL (\MXXL) $N$-body simulation of \cite{Angulo2012} by
\cite{Smith:2017tzz}. This mock is tuned match the luminosity function, colour distribution, and clustering properties of the \SDSS \MGS at low redshift, and the  evolution of these statistics to redshift $z \approx 0.5$ as measured from the \GAMA survey \citep{2012yCat..74130971D, 2015MNRAS.452.2087L, 10.1093/mnras/stx3042}.

The \DESI \BGS is expected to have a target density of just over $800$ galaxies per square degree in a primary sample defined by a faint $r$-band magnitude limit of $19.5$. Then, in a lower priority sample, a secondary sample of $\sim600$~ galaxies $\textrm{deg}^{-2}$ defined by the magnitude range $19.5 < r < 20$ \citep{DESI2016:surveys}. From hereon in we will refer to these \BGS samples as \BGSB and \BGSF respectively. A few per cent of galaxies in the \DESI \BGS will be lost due to deblending errors, superposition with bright stars, and other artifacts that typically affect imaging catalogues. Our aim is to provide a reliable input galaxy catalogue for the \DESI \BGS and to characterize its properties, such as the surface density of galaxies and their clustering. A complementary study by \cite{10.1093/mnras/staa1621} examined the impact of imaging systematics on the selection and clustering of targets in the LRG, ELG and QSO DESI surveys, using an earlier release of the Legacy Surveys imaging data \citep{Dey:2019}.

Here, we define and characterized the \BGS target selection based on the latest \DECaLS release, DR8, which covers $\sim 2/3$ of the full $14\,000$ $\textrm{deg}^2$ of \DESI footprint. 
The resulting catalogue is defined in
\cite{Ruiz_Macias_2020} and here we present the details of that selection and associated analysis of the catalogue.
This \BGS catalogue  was used by \DESI in the commissioning stage of the early survey validation observations. It is planned that the final BGS catalogue will be based on the next, DR9, Legacy Survey data release. This release will include better modelling 
of large galaxies and the light in bright star haloes.
More discussion of DR9 and planned subsequent characterization of the \BGS selection   can be found in Section~\ref{sec:conclusions}.

This paper is organised as follows: in Section~\ref{sec:data_sets} we describe the Legacy Surveys imaging data used to select our targets and the secondary datasets used to tune the selection. In Sections~\ref{sec:spatial_masking} and~\ref{sec:photo_select} we define the spatial and photometric cuts used to select \BGS targets and to get rid of artifacts that might become problematic for \DESI observations plus the removal of poor quality imaging data. In Section~\ref{sec:photo_select} we define our star-galaxy classification using \GAIA DR2. In Section~\ref{sec:cat_properties} we compare the \BGS catalogue with its overlap of the
\GAMA DR4\footnote{This is an unreleased version of \GAMA catalogue that the \GAMA collaboration made available to us. It is essentially the same as \GAMA DR3, but with more redshifts.} \citep{2012yCat..74130971D, 2015MNRAS.452.2087L, 10.1093/mnras/stx3042} to assess the completeness and contamination of the \BGS and to quantify its expected redshift distribution.
In Section~\ref{subsec:systematics} we look at eight potential systematics that might be affecting our \BGS target selection and try to mitigate these effects with linear weights determined using the stellar density. Section~\ref{subsec:angular_correlation} shows the clustering of our \BGS selection before and after applying the weights and we compare it with \SDSS and the \MXXL lightcone catalogue \citep{Smith:2017tzz}. Finally, in Section~\ref{sec:conclusions}, we summarize our results and present our conclusions.

\section{Photometric Data sets}\label{sec:data_sets}

During the \BGS target selection process we make use of several catalogues. The main data set used is the Legacy Surveys \DReight (hereafter \LS \DReight) imaging catalogue from which we select our targets. We also make use of secondary catalogues for masking purposes, such as the \TYCHO star catalogue \citep{2000A&A...355L..27H}, 
 the \GAIA DR2 \citep{2016A&A...595A...1G},
the Siena Galaxy Atlas - 2020 (\SGA-2020) (Moustakas in prep.) and globular clusters from the OpenNGC\footnote{OpenNGC, \url{https://github.com/mattiaverga/OpenNGC}, 
is a database containing positions and main data of NGC (New General Catalogue) and IC (Index Catalogue) objects constructed by the GAVO data center team by merging data from NED, HyperLEDA, SIMBAD, and several databases available at HEASARC (\url{https://heasarc.gsfc.nasa.gov/}).} catalogue. We also use a combination of \GAIA DR2 and \LS photometry to perform star-galaxy separation.

\subsection{Legacy Survey DR8 (DECam)}\label{subsec:lsdr8}

The Dark Energy Camera Legacy Survey (\DECaLS), the Beijing-Arizona Sky Survey (\BASS), and the Mayall $z$-band Legacy Survey (\MzLS) together constitute the \DESI Legacy Imaging Survey (hereafter the Legacy Survey). The imaging Legacy Survey was created with the aim of attaining photometry with the necessary target density, coverage and depth required for \DESI . The \SDSS \MGS  \citep{2002AJ....124.1810S} and Pan-STARRS1 \citep{2016arXiv161205560C} 
catalogues are both too shallow to be used to reliably select the \DESI survey targets. The \DES survey \citep{DES:2005} does reach the target depth for \DESI, but only covers $5000$ deg$^2$, mostly in the South Galactic Cap (\SGC), with only $\sim1130$ deg$^2$ observable with \DESI.

This work is based on the eighth release of the Legacy Survey project (LS \DReight) which is the first release to  integrate data from all of the individual components of the Legacy Surveys (\BASS, \DECaLS and \MzLS). However, this paper focuses only on \DECaLS data.

\begin{figure*}
	\includegraphics[width=16cm]{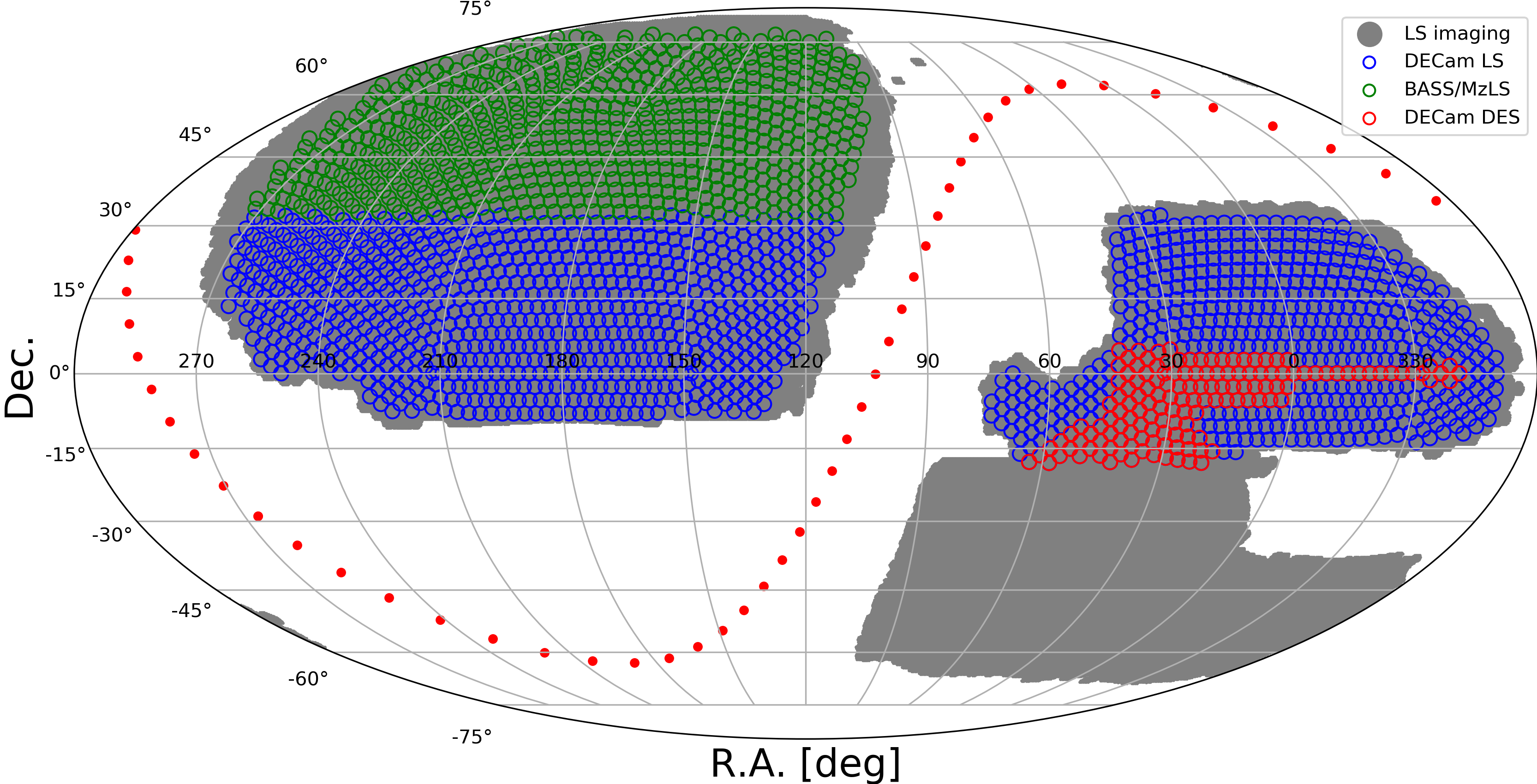}
    \caption{The sky map of the footprint of all the \LS imaging used in \DECaLS and in \BASS and \MzLS is shown in gray. The red and blue circles show the \DESI tiles that define the portion of \DESI survey footprint that lies within \DECaLS.
    The blue tiles are those for which the data comes from the \DECam \LS  imaging while the red tiles come from \DECam \DES imaging. The green tiles show the northern \DESI footprint whose imaging data comes from the BASS and MzLS surveys which are not the focus of this paper. The red dots show the locus of the Galactic plane.}
    \label{fig:decals_in_desi}
\end{figure*}

The \DECaLS data in the \LS \DReight data release comprises observations from $9$th August $2014$ through $7$th March $2019$. \DECam images come from the Dark Energy Camera \citep[\DECam][]{2015AJ....150..150F} at the $4$-m Blanco telescope at the Cerro Tololo Inter-American Observatory. \DECam has $62$ $2048\times 4096$ pixel format $250 \mu$m-thick \LBNL \CCDs arranged in a roughly hexagonal $\sim3.2$ deg$^2$ field of view. The pixel scale is $0.262$ arcsec/pix and the camera has high sensitivity across a broad wavelength range of $\sim400-1000$~nm. 
Since LS \DReight data goes beyond the intended \DESI footprint\footnote{Current LS \DReight imaging covers around $\sim20\,332$ deg$^2$ of which $15\,174$ deg$^2$ corresponds to \DECaLS.}  of $\sim14\,000$ deg$^2$, we are going to consider only data within the \DESI footprint. This corresponds to $\sim9\,717$ deg$^2$ of \DECaLS data of which $\sim1\,114$ deg$^2$ are covered by \DECam data coming from the \DES \citep{DES:2005}. We essentially have two \DECam data sets, i) \DECam imaging taken for the \LS programme which we refer to as \DECam \LS  and ii) the \DECam data coming from the \DES programme which we refer to as \DECam \DES. \DECam \LS and \DECam \DES combine to form the \DECaLS data set. Fig.~\ref{fig:decals_in_desi} shows the sky map coverage of \DECaLS imaging indicating the  \DECaLS imaging that lies within the \DESI footprint. \DECaLS is the only survey that covers the entire \SGC ($4\,394$ deg$^2$) and the \NGC ($5\,323$ deg$^2$) regions of the \DESI survey at declination $\delta \leq +\ang{32.375}$. 

In order to fulfil the target selection required for the different \DESI surveys (\BGS, \LRGs, \ELGs and \QSOs), it was concluded that a three-band $g$, $r$ and~$z$ optical imaging programme, complemented by {\it Wide-field Infrared Survey Explorer} ({\it WISE}) \Wone and \Wtwo photometry, would be sufficient. The minimal depth\footnote{The depths are defined as the optimal-extraction (forced-photometry) depths for a galaxy near the limiting depth of \DESI, where that galaxy is defined to be an exponential profile with a half-light radius of $r_{\rm half}= 0.45$~arcsec.} required is $g=24.0$, $r=23.4$ and $z=22.5$. \DECam  \LS reaches these required depths in total exposure times of $140$, $100$ and $200$ sec in $g$, $r$, $z$ respectively in nominal\footnote{Here `nominal' is defined as photometric and clear skies with seeing FWHM of $1.3$~arcsec, airmass of $1.0$, and sky brightness 
in $g$, $r$ and $z$ of $22.04$, $20.91$ and $18.46$ AB mag arcsec$^{-2}$, respectively.} conditions, typically in a minimum of two visits per field.

All data from the Legacy Surveys are first processed at the NSF’s National Optical-Infrared Astronomy Research Laboratory in Tucson (\NOAO) through the \NOAO Community Pipeline\footnote{\url{https://www.noao.edu/noao/staff/fvaldes/CPDocPrelim/PL201\_3.html}} (\CP). The \CP takes raw data as an input and provides detrended and calibrated data products such as instrumental calibration (e.g. bias subtraction and flat fielding), astrometric calibration (e.g. mapping the distortions and providing a world coordinate system, or \WCS), photometric characterization (e.g. magnitude zero point calibration) and artifact identification, masking and/or removal (e.g. removal of cross-talk and pupil ghosts, and identification and masking of cosmic rays). 

The source catalogues for the Legacy Surveys are constructed using the \LEGACYPIPE\footnote{\url{https://github.com/legacysurvey/legacypipe}} software, which uses the \TRACTOR \footnote{ \url{https://github.com/dstndstn/tractor}}\citep{2016ascl.soft04008L} code for pixel-level forward-modelling of astronomical sources.  This is a statistically rigorous approach to fitting the differing point spread functions (\PSF) and pixel sampling of these data, which is particularly important as the optical data have a typical \PSF width of $\sim1$~arcsec.

The steps in the \LEGACYPIPE processing are described in \cite{Dey:2019}; we briefly summarize relevant parts here.

After initial source detection
and defining the contiguous set of pixels associated with each detection (termed a blob), \LEGACYPIPE proceeds to fit these pixels with models of the surface brightness, including a point-source and a variety of galaxy models.  These fits are performed on 
the individual optical images (in $g$, $r$ and $z$ bands), taking into account the different \PSF and sensitivity of each image, using \TRACTOR.

Besides the \PSF model, \TRACTOR fits four other light profile models to sources: a round exponential with a variable radius (referred to as \REX), an exponential profile (\EXP), a de Vaucouleurs profile(\DEV), and a composite of \DEV and \EXP profiles (COMP). The decision as to whether or not to retain an object in the catalogue and the choice of the model to best describe its light profile is treated as a penalized-$\chi^{2}$ model selection problem.

This process results in object fluxes and colours that are consistently measured across the wide-area imaging surveys that form the input into the \DESI target selection. In general, \TRACTOR improves the target selection for all \DESI surveys by allowing information from low resolution and low signal-to-noise measurements to be combined with those from high resolution and high signal-to-noise data. The \TRACTOR catalogues include source positions, fluxes, shape parameters, and morphological quantities that can be used to discriminate extended sources from point-sources, together with errors on these quantities.  The \BGS is flux limited in the $r$-band. However, since \TRACTOR performs simultaneous fits in $g$, $r$ and $z$ we also chose to impose quality cuts in the other bands as well as those in the $r$ band when selecting the \BGS targets. 

The main \TRACTOR outputs required for the \BGS are the total fluxes\footnote{The fluxes output by \TRACTOR are in units called \NANOMAGIES. A flux of $1$ \NANOMAGIE corresponds to an AB magnitude of $22.5$.} corresponding to the best-fitting source model (i.e., \PSF, \REX, \EXP, \DEV or \COMP) in all three bands ($g,r$ and $z$), the number of observations (\NOBS) in the three bands, the predicted flux (in the $r$-band only) within the aperture of a fibre which is around $1.5$ arcsec diameter (\FIBERFLUX\footnote{The \FIBERFLUX is in units of \NANOMAGIES}) in $1$~arcsec Gaussian seeing. The Galactic extinction values are derived from the SFD98 maps \citep{1998ApJ...500..525S} and are reported in linear units of transmission (\MWTRANSMISSION) in the $g,r$ and $z$ bands, with a value of unity representing a fully transparent region of the Milky Way and $0$ indicating a fully opaque region. The extinction coefficients for the \DECam filters were computed through an airmass of $1.3$, for a source with a $7\,000$~K thermal  spectrum \citep{2011ApJ...737..103S}. The resulting coefficients are $A / E(B-V) = 3.995$, $3.214$, $2.165$, $1.592$, $1.211$, $1.064$ in $ugrizY$. These are then multiplied by the SFD98 $E(B-V)$ values at the coordinates of each object to derive the $g, r$ and $z$ \MWTRANSMISSION values. Finally, in each band, there is a set of quality measures called \FRACMASKED, \FRACFLUX and \FRACIN that quantify the quality of the data in each profile fit. We describe these in more detail in Section~\ref{sec:QC}.

The fluxes returned by \TRACTOR can be transformed into AB magnitudes as follows:
\begin{eqnarray}  
    magr  \negthinspace \negthinspace \negthinspace  \negthinspace  &=& \negthinspace  \negthinspace  \negthinspace  \negthinspace 22.5-2.5 \,\log_{10}(\rm \FLUX) , \label{eq:mag_uncorr} \\
    mag^{\hphantom{*}}\negthinspace \negthinspace  \negthinspace  \negthinspace  &=& \negthinspace  \negthinspace  \negthinspace 
    \negthinspace 
    22.5 -2.5 \, \log_{10}(\rm \CORRFLUX) , \label{eq:mag_corr}
\end{eqnarray}
where Eqn.~(\ref{eq:mag_uncorr}) does not include the correction for Galactic extinction, unlike Eqn.~(\ref{eq:mag_corr}). The $r$ in Eqn.~(\ref{eq:mag_uncorr}) stands for {\it raw}.

Table~\ref{tab:band_passes} shows the area covered by photometry in each of the three bands of \DECaLS \DReight with $1$, $2$ or $3$ passes.
These values are just for the data within the \DESI footprint, as shown in Fig.~\ref{fig:decals_in_desi}. This \DECaLS footprint covers a total of $9\,717$ deg$^2$. Expressed in percentages, $99.5 \,$ per cent of this area has at least one pass in all of the three bands 
$grz$, $95.3 \,$ per cent has at least two passes and $70.7 \,$ per cent has at least three passes in all three bands.

\begin{table}
\caption{The area, in square degrees, of \DECaLS \DReight covered by at least $1$, $2$ or $3$~passes in each of the three filters ($grz$) individually (first three rows), and combined (i.e. at least $1$, $2$ or  $3$~passes in each of the $3$ bands; bottom row).
We have restricted our results to observations within the \DESI footprint as shown in  Fig.~\ref{fig:decals_in_desi}.}
\label{tab:band_passes}
\centering
\begin{tabular}{ |p{2.9cm}||p{1.2cm}|p{1.2cm}|p{1.2cm}|  }
 \hline
 {\bf Band/Number of Passes} & $\geq 1$ & $\geq 2$ & $\geq 3$\\
 \hline
 $g$-band   & $9\,687$  & $9\,454$  & $7\,769$  \\
 $r$-band   & $9\,686$  & $9\,422$  & $7\,569$  \\
 $z$-band   & $9\,686$  & $9\,487$  & $8\,036$  \\
 combined & $9\,669$    & $9\,257$  & $6\,870$  \\
 \hline
\end{tabular}
\end{table}

\subsection{Secondary catalogues}

Here we list other catalogues that are used either to exclude regions of the sky in which the extraction of galactic sources is compromised by the presence of other objects, or to perform star-galaxy separation. 

\subsubsection{Tycho 2}

Bright stars can impinge upon the estimation of the photometric properties of nearby galaxies or may even lead to the generation of spurious sources. Hence, it is prudent to simply exclude or veto regions close to known bright stars to avoid such problems. 
Regions near bright stars are masked out of the target catalogue using the \Tycho catalogue \citep{2000A&A...355L..27H}. The \Tycho catalogue contains positions, proper motions, and two-colour photometry for $2\,539\,913$ of the brightest stars in the Milky Way.

\subsubsection{Gaia DR2}\label{subsec:gaia}

\GAIA \citep{2016A&A...595A...1G} is a European Space Agency mission that was launched in 2013 with the aim of observing $\approx 1$ per cent of all the stars in the Milky Way, measuring accurate positions for them along with their proper motions, radial velocities, and optical spectrophotometry. The wavelength coverage of the astrometric instrument, defined by the white-light photometric $G$-band magnitude, is $330$ -  $1050$~nm \citep{2016A&A...595A...7C}. These photometric data have a high signal-to-noise ratio and are particularly  suitable for variability  studies.

Since the first release of \GAIA data \citep{2016A&A...595A...2G}, this survey has been widely used by the \DESI \LS (i.e. for astrometric calibrations, proper motions, bright star masking) and is also ideal for constructing a star-galaxy separator for the \BGS. There are $1.7$ billion stars in the second \GAIA data release (DR2)\footnote{DR2 covers 22 months of observations and was released on 25 April 2018.}, over the whole sky to $G = 20.7$, which is sufficiently deep to detect all stars that might contaminate the \BGSF sample. We describe how we use a combination of \GAIA and \LS photometry to perform star-galaxy separation in Section~\ref{subsec:star_galaxy}.

\subsubsection{Globular clusters and planetary nebulae}

Globular clusters and planetary nebulae are bright extended sources that can affect the identification of extragalactic sources in a similar way to bright stars. In the \LS, an area of sky around such objects is excluded to minimize their impact on target selection. The Open\NGC catalogue\footnote{\url{https://github.com/mattiaverga/OpenNGC}} is used to provide a list of such sources. The extent and impact of masking around globular clusters and planetary nebulae is discussed in Section~3.1.3.

\subsubsection{The Siena Galaxy Atlas}\label{subsec:SGA}

Large galaxy images can be broken up by photometric pipelines, which, for example, could mistake H\,\textsc{II} regions inside the galaxy for individual extended sources. Also, spurious sources could be generated around the boundaries of large galaxies. 
The Siena Galaxy Atlas - 2020 (\SGA-2020)\footnote{\url{https://github.com/moustakas/SGA}} is an ongoing project  to select the largest galaxies in the \LS using optical data from the HyperLeda catalogue\footnote{\url{http://leda.univ-lyon1.fr/}}  \citep{2014A&A...570A..13M} and infrared data from the \ALLWISE catalogue \citep{2015ApJS..221...12S}. Currently the catalogue contains $535\,292$ galaxies that have an angular major axis (at the $25$~mag/arcsec$^2$ isophote) larger than $20$~arcsec. The use of the \SGA-2020 in the spatial mask of the \BGS is described in Section~3.1.2. 

\section{Spatial Masking}\label{sec:spatial_masking}

Our main goal is to produce a reliable \BGS input catalogue that fulfils the \DESI science requirements. 
If the target list contains spurious objects, these will mistakenly be allocated fibres leading to a reduction in the efficiency and completeness of the redshift survey. Furthermore, spurious objects could imprint a systematic effect in the measured clustering.  

A step towards minimising the number of spurious objects  is to mask out regions of the sky around bright stars, since features such as extended halos, ghosts, bleed trails and diffraction spikes around the stars can compromise the measurement of the photometry of neighbouring objects. Similarly we must remove areas around very large galaxies and globular clusters and planetary nebulae; such objects can also affect the photometric measurements of their neighbours, leading to incorrect properties or spurious objects.  

Within the same framework, we have to propagate  instrumental effects such as saturated pixels, bad pixels, bleed trails, etc. that the \NOAO \CP tracks and \TRACTOR reports in the \LS catalogue\footnote{In the \LS \DReight catalogue information on whether or not the photometric parameters measured for an object have the possibility of being influenced by a bad pixel is flagged by the \ALLMASK \MASKBITS.}

One way to avoid contamination of the catalogue with spurious objects is to exclude regions around bright stars and galaxies. This can be done with a simple but effective circular mask for stars and by using elliptical masks for galaxies. In Section~\ref{subsec:geometric} we set out the geometrical masking functions we have applied around bright stars, large galaxies and globular clusters to minimize the number of spurious targets in our \BGS catalogue.  In Section~\ref{subsec:pix_masking} we describe the masks applied to reduce the number of spurious targets due to imaging artifacts such as bad pixels resulting from saturation and bleed trails. 

For subsequent analysis (e.g. estimating clustering statistics), it is very important to keep a record of the areas of the survey that are removed by these masks. For this purpose we have made use of the randoms catalogue developed by the \DESITARGET\footnote{\url{https://github.com/desihub/desitarget}} team. The randoms catalogue has a total density of $50\,000$ objects/deg$^{2}$ divided into $10$ subsets, each with density of $5\,000$ objects/deg$^{2}$. Each random carries with it some of the \DECam imaging information computed from the image pixel (in each band and exposure) in which it is located and supplementary information such as the dust extinction extracted from HEALPix\footnote{\url{http://healpix.sourceforge.net}} maps \citep{Zonca2019}. These imaging attributes include the number of observations (\NOBSG, \NOBSR, \NOBSZ), galactic extinction (\EBV), the bitwise mask for optical data (\MASKBITS), etc\footnote{For more information on the properties of randoms see: \url{http://legacysurvey.org/dr8/files/\#random\-catalogs}}.

\begin{figure}
	\includegraphics[width=\columnwidth]{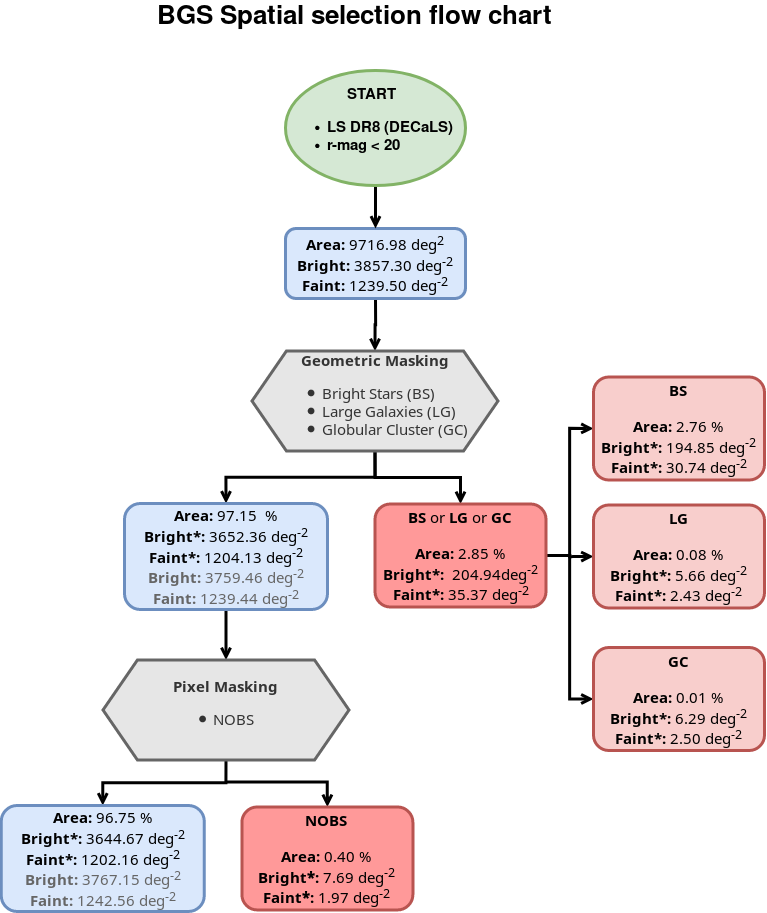}
    \caption{The flow chart shows the effects of the spatial masks that are applied as part of \BGS target selection for the \DECaLS \DReight data. The spatial masking is divided into two classes, one defined by the geometrical cuts which exclude regions around bright sources (bright stars, large galaxies and globular clusters), and the other by pixel-based cuts which use information such as the number of observations (\NOBS). The boxes in the flow chart show the survey area (in deg$^2$) and the target number density (per square deg) split into \BGSB and \BGSF after each mask is applied. The blue boxes give this information for the portion of the survey that is retained while the red boxes give this information for the areas removed.
    If more than one mask is combined at a single stage (as indicated within the gray hexagonal boxes), then the dark-red boxes show the results for the combination of these masks and the light-red boxes shows the results for each individual mask. As some of the masks can overlap the numbers in the light-red boxes do not necessarily add up to those in the dark-red boxes.
    The target densities with the ($^*$) superscript are computed without correcting for the area removed by the masking while those without the ($^*$) superscript are corrected for the masked area.
    The gray hexagonal boxes describe the different masks. Note that star-galaxy separation is not yet applied here and this is why we have a high target density in the blue boxes.}
    \label{fig:flow1}
\end{figure}

In Fig.~\ref{fig:flow1} we show a flow chart which summarizes the spatial masking applied when constructing the \BGS catalogue. The spatial masking is broken down into two classes: {\it geometrical masking} and {\it pixel masking}. The blue boxes of the flow chart report the survey area (in deg$^2$) and mean target densities (in objects/deg$^{2}$) after successively applying each mask (gray hexagonal boxes). The red boxes record the same information for the rejected area and objects. The final \BGS catalogue does not depend on the order in which the masks are applied, but as some areas and targets are rejected by more than one mask the information in the red boxes depends on the ordering. For example, the area and number of objects shown as being rejected by the pixel masking excludes what would be rejected by this mask if the geometric masks had not been applied first. Overall, for the \DECaLS footprint of $9\,717$ deg$^2$, the spatial masking removes $3.25$ per cent of the area.

\subsection{Geometrical  masking}\label{subsec:geometric}
\subsubsection{Bright star mask (\BS)}\label{subsubsec:BS}

The bright star (\BS) mask is based on the locations  of stars from \GAIA DR2 \citep{2018A&A...616A...1G} and the \TYCHO \citep{2000A&A...355L..27H} catalogue  after correcting for epoch and proper motions.
This mask
consists of the union of circular exclusion regions around each
star, where the radius of the exclusion region, estimated from an earlier stacking analysis, 
depends on the magnitude of the star in the following way:

\begin{eqnarray}
\label{eq:mag_radii_bs}
  R_{\rm BS}(m)  &=&  
  39.3 \times 2.5^{(11 - m)/3} \,\, {\rm arcsec},   \hspace{0.2cm} m  > 2.9 \\ \nonumber 
 &= &      471.6 \,\, {\rm arcsec}, \hspace{1.85cm} m  < 2.9.
\end{eqnarray}
Here $m$ is either \GAIA $G$-mag or \Tycho \textsc{mag}\_\textsc{vt}  with \GAIA $G$-mag being used when both are available.  Stars fainter than $m=13$ have no exclusion zone around them.  

The \BS masking uses a total of $773\,673$ \GAIA DR2 objects ($82$ objects/deg$^{2}$) with \GAIA $G$-mag brighter than $13$, while from \Tycho, we have a total of $3\,349$ objects ($\sim0.36$ objects/deg$^{2}$) to a \TYCHO visual magnitude brighter than  \textsc{mag}\_\textsc{vt} $=13$. In order to avoid overlaps both catalogues have been matched after applying proper motions to bring \GAIA objects to the same epoch as \Tycho and keeping only the \Tycho objects that are not found in \GAIA. These \Tycho stars represents only a $0.4 \%$ of total stars used for the \BS masking. Then the magnitude, $m$, used to compute the mask radius in equation~(\ref{eq:mag_radii_bs}) is the \GAIA $G$-band magnitude
for the \GAIA stars and the \Tycho visual magnitude, \textsc{mag}\_\textsc{vt}, for the retained \Tycho stars. 
The overall median difference between the \Tycho and \GAIA   
magnitude is $0.4$ with \Tycho being fainter. This $0.4$ magnitude difference translates into a median decrease in masking radius of $50$~arcsecs for \GAIA stars with magnitude of $3$ and a decrease of $2$~arcsecs for \GAIA stars with magnitude of $13$ from equation \ref{eq:mag_radii_bs}.
Within $R_{\rm BS}(m)$ \TRACTOR forces all the sources it detects to be fit with the \PSF profile to avoid artificially fitting diffraction spikes and stellar haloes as large extended sources. Thus any galaxies detected within $R_{\rm BS}$ will have their fluxes underestimated. Consequently to define a reliable galaxy catalogue we must veto all sources within  $R_{\rm BS}$ of a bright star. In Fig.~\ref{fig:flow1} we show that this Bright star mask covers $2.76$ per cent of the initial footprint and rejects $\sim$$195$ potential \BGSB  objects/deg$^{2}$ and $\sim$$31$ potential \BGSF  objects/deg$^{2}$ when averaged over the full initial footprint. It should be noted that most of these objects are stars as star-galaxy separation has not been applied at this stage in the flow chart shown in Fig.~\ref{fig:flow1}.
An alternative ordering of the flow chart with star-galaxy separation applied first is shown in Fig.~\ref{fig:flow_galaxy}. 
There we see that for galaxies the corresponding numbers are $13.7$ galaxies/deg$^{2}$ 
for \BGSB  and $8.5$  galaxies/deg$^{2}$ for \BGSF. 


\begin{figure*}
	\includegraphics[width=16cm]{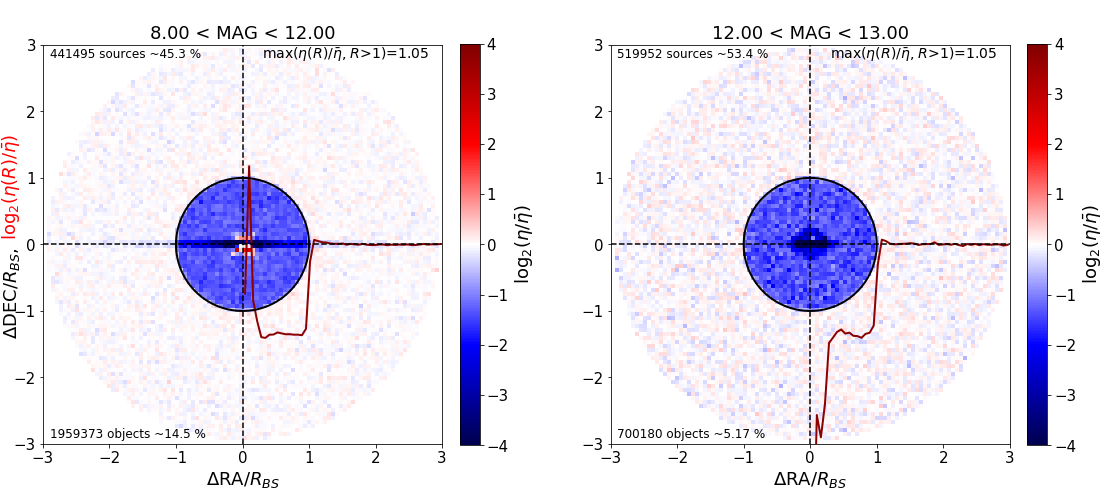}
    \caption{2D histograms of the positions of \BGS objects  relative to their nearest Bright Star (\BS) taken from the \GAIA and \Tycho sources down to $G$-mag and visual magnitude \textsc{mag}\_\textsc{vt} of $13$ respectively. 
    These stacks are performed in magnitude bins in the \BS catalogue from magnitude $8$ to $12$ (left) and $12$ to $13$ (right). The stacks are made using angular separations rescaled to the masking radius function given in Eqn~\ref{eq:mag_radii_bs}, which means that objects within a scaled radius of $0$ to  $1$ will be masked out by the \BS veto while objects with $R = r/R_{\rm BS} > 1$ will not (here $r^2 = (\Delta {\rm RA}^2 \cos({\rm DEC})^2 + \Delta {\rm DEC}^2$).
    The colour scale shows the ratio of the density per pixel ($\eta$) to the mean density ($\bar{\eta}$) within the shell $1.1 < r/R_{\rm BS} < 3$. The density ratio is shown on a $\log_2$ scale where red shows overdensities, blue corresponds to  underdensities and white shows the mean density. The black solid circle shows extent of the \BS exclusion zone. The red solid line shows the radial density profile on the same scale as the colour distribution $\log_2(\eta(R)/\bar{\eta})$ where $\eta(R)$ is the target density within the annulus at radius $R$ of width $\Delta R \sim 0.06$.}
    \label{fig:2dstack_tycho_rescaled}
\end{figure*}

To determine if the bright star mask is adequate or whether the effects of stellar haloes causes a systematic error in the photometry of neighbouring galaxies that extends to larger radii, we plot in Fig.~\ref{fig:2dstack_tycho_rescaled} the average density of \BGS galaxies in the
vicinity of bright stars prior to applying the bright star mask. If the photometry of galaxies has been compromised in any means, this can be seen in the galaxy number density to a fixed magnitude due to the strong dependence of galaxy number density on apparent magnitude. The term \BGS galaxy refers to the BGS sample after applying the star-galaxy separation and the spatial and photometric cuts down to the $r$-band magnitude of $20$, which will be covered in the subsequent subsections of Section~\ref{sec:spatial_masking} and in Section~\ref{sec:photo_select}. The stacks are made by expressing the angular separation, $r$, of the \BGS galaxies prior to apply the bright star mask from their nearest bright star in units of the bright star masking radius $R_{\rm BS}$, as given by Eqn.~\ref{eq:mag_radii_bs}. In these rescaled coordinates, $R = r/R_{\rm BS}$, galaxies within a radius of unity, shown by the black circle, are within the BS masking zone. We show stacks for two magnitude bins defined by the $G$-mag and visual magnitude \textsc{mag}\_\textsc{vt} for \GAIA \DRtwo and \Tycho stars respectively, one with bright stars of magnitude between $8$ to $12$ and one fainter with magnitude between $12$ to $13$. The radial profile (red solid line) shows the variation in the target density, defined as $\Delta\rho(R) \equiv\log_2({\eta}(R)/\bar\eta)$ where ${\eta}(R)$ is the target density in an annulus at radius $R$ of width $\Delta R \sim 0.06$, and $\bar\eta$ is the mean target density evaluated over the region $1.1 < R < 3$. This means that $\Delta\rho(R) = 0$ corresponds to the mean density, $\Delta\rho(R) \ge 1$ to an  overdensity at least twice the mean density, and $\Delta\rho(R) < 0$ to an  underdensity. The large underdensity at radius $R \le1$ is due to \TRACTOR forcing all objects within this region to be fit by the \PSF model. In Section~\ref{subsec:star_galaxy} we will see how stars and galaxies are defined for \BGS target selection, which does not depend on \TRACTOR \PSF designation, therefore, galaxies in the region $R < 1$ are allowed. In the left panel of  Fig.~\ref{fig:2dstack_tycho_rescaled}, we see a spike of spurious galaxies for $R < 0.2$. In contrast the right panel shows a strong deficit of galaxies at $R < 0.2$. For $R > 1$, the stacks show uniform density close to mean, suggesting the star mask is working. There is a small bump just outside the masking radius where a $\sim6$ per cent excess is seen in both panels. This may need to be revisited for accurate clustering studies, 
but is not large enough to be a concern for the efficiency of target selection.

\subsubsection{Large galaxies mask (\LG)}
\label{subsubsec:LG}

Without special treatment, large galaxies in which spiral arms and other structures such as H\,II regions are resolved would be artificially fragmented by \TRACTOR into multiple sources. To avoid this and to achieve more accurate photometry  for large galaxies in the \SGA-2020 catalogue (see \S\ref{subsec:SGA}), \TRACTOR is seeded with different priors, and within an elliptical mask centred on the large galaxy \TRACTOR fits secondary detections using only the \PSF model. This reduces the spurious fragmentation of large galaxy images, but also means that genuine neighbouring galaxies within the masked area have compromised photometry. The elliptical mask that is used has the same position, $25$~mag/arcsec$^2$ isophotal major axis angular diameter, D$_{25}$, semi-minor to semi-major ratio, $B/A$ and position angle, $PA$ as the ones used to define the large galaxies in the \SGA-2020 catalogue.  Defining an effective masking radius of
$r=\sqrt{ab}$, where $a$ and $b$ are the semi-major and semi-minor axes of the elliptical mask, the median masking radius for the \LG galaxies is $10.8$~arcsecs.

We apply these same masks to reject objects from the \BGS catalogue but then we reinstate the large galaxies provided they are not also masked by the bright star or globular cluster mask. 
The area covered by the combined \LG mask amounts to only $0.08$ per cent of the initial area and the number of objects removed amounts to $~5.7$  objects/deg$^{2}$ \BGSB and $~$$2.4$  objects/deg$^{2}$ \BGSF objects over the full initial area.

\subsubsection{Globular cluster mask (\GC)}

The globular cluster (\GC) mask works in a similar way to the \BS mask, by applying a circular exclusion zone around the \GC. The masking radius is defined by the major axis attribute for the object in the OpenNGC catalogue.

The GC mask has the smallest impact of the geometric masks, rejecting only $0.01$ per cent of the initial area, accounting for densities of $~6.3$  objects/deg$^{2}$ in \BGSB and $~2.5$  objects/deg$^{2}$ in \BGSF. \TRACTOR also force fits as \PSFs everything within this mask.

\subsection{Pixel masking}\label{subsec:pix_masking}
Some of the effects that compromise the photometry on a pixel basis and the model fitting include bad pixels, saturation, cosmic rays, bleed trails, transients. The \NOAO \DECam \CP identifies these instrumental effects during its various calibrations\footnote{The document that lists all the calibrations and which includes details about the various maskings can be found at: \url{https://www.noao.edu/noao/staff/fvaldes/CPDocPrelim/PL201\_3.html}} (see Table $5$ in \cite{Dey:2019} for a list of the calibrations) and these are passed through \TRACTOR and compiled in the \ALLMASK \BITMASK\footnote{Details of this \BITMASK can be found here: \url{http://www.legacysurvey.org/dr8/bitmasks/\#allmask-x-anymask-x}}. \ALLMASK denotes a source blob that overlaps with any of the mentioned bad pixels in all of the overlapping images.

Besides the bad pixels which arise due to instrumental defects, the \BGS requires a complete sample in the three bands ($g, r$ and $z$). We therefore impose a requirement that there is at least one observation in each of the bands through the \NOBS parameter. \NOBS stands for Number of Observations, and is defined as the number of images that contributes to the source detected central pixel in each of the bands.  Both \ALLMASK and \NOBS are pixel-based and hence this information is also available in the random catalogue. However, we find that virtually all of the area ($~97$ per cent) (and hence virtually all of the randoms) rejected by \ALLMASK is also rejected by using \NOBS$=0$ (in any band). In addition, \ALLMASK rejects a  significant number of objects ($196$ objects/deg$^{2}$) but with a small associated area ($~0.01$ per cent of the full area). Virtually all the objects rejected by \ALLMASK and many others are already rejected by the quality cuts in \FRACMASKED, \FRACIN and \FRACFLUX (in any band); these cuts will be reviewed in Section \ref{sec:photo_select}. 

In conclusion, there is little to be gained from using \ALLMASK and we have therefore decided to use only \NOBS as our pixel level mask,  shrinking the area by $0.4$ per cent and reducing the target density by $7.7$ objects/deg$^{2}$ in \BGSB and $2$  objects/deg$^{2}$ in \BGSF.

\section{Photometric selection}\label{sec:photo_select}

Following the spatial masking described in the previous section, the next step in the construction of the \BGS target list is to incorporate information about photometric measurements into the selection process.  According to the science requirements of the \BGS and the mock \BGS catalogues made by \cite{Smith:2017tzz}, the survey is expected to have a target density of $~800$ galaxies deg$^{-2}$ to an $r$-band limit of 19.5. For the faint sample (19.5 < $r$ < 20), which is second priority in \BGS, a density of $~600$ galaxies deg$^{-2}$ is expected.

\begin{figure}
	\includegraphics[width=\columnwidth]{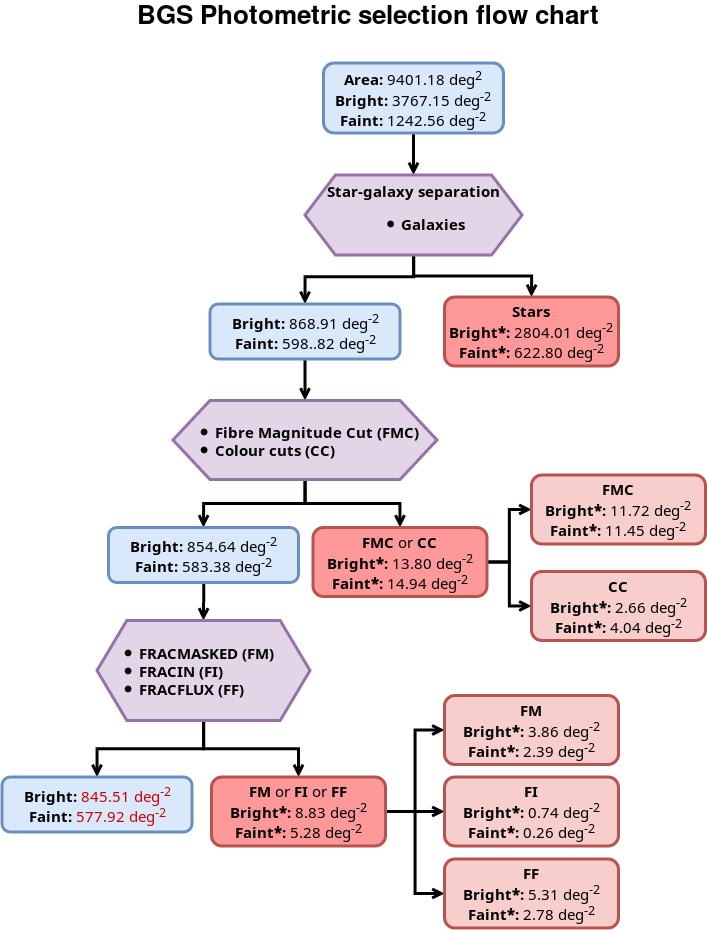}
    \caption{Flow chart of the \BGS target selection in the Legacy Surveys \DReight based on photometric considerations. The photometric selection of \BGS targets is divided into four stages; star-galaxy separation, fibre magnitude cuts (\FMC), colour cuts (\CC) and quality cuts (\QCs). The photometric cut flow chart is a continuation of the spatial cut flow chart (Fig.~\ref{fig:flow1}) and therefore we start from the area and object densities reported at the end of the spatial cut flow chart. We report densities for the bright and faint samples separately, showing in blue boxes the values for the sources remaining after each of the \BGS cuts. The densities of the removed objects are shown in red/pink boxes. The different cuts applied are shown in purple hexagonal boxes.}
    \label{fig:flow2}
\end{figure}

One of the major challenges for the \BGS is the separation of stars and galaxies. In Section~\ref{subsec:star_galaxy} we describe how we
compare high angular resolution point source magnitudes from
\GAIA DR2 \citep{2018A&A...616A...1G} 
with total magnitudes from the best-fitting light profile model selected by \TRACTOR to distinguish point sources from extended sources. 

In Section~\ref{subsec:fibrecut} we describe how we reject spurious objects that have incongruous light profiles by comparing their total 
magnitudes with the fibre magnitude that \TRACTOR computes from the fitted profile assuming $1$~arcsec Gaussian seeing and $1.5$~arcsec fibre diameter. We place a cut in the fibre magnitude versus total magnitude plane that is motivated by the locus of confirmed galaxies from the  \GAMA DR4 survey.

Further posterior cuts which use photometry include removing colour outliers in $g-r$ and $r-z$ (see \S~\ref{subsec:colourcuts}), and applying quality cuts that indicate low accuracy in the flux measurement for an object (see \S~\ref{subsec:fraccuts}). The quality cuts make use of the quantities \FRACMASKED, \FRACFLUX and \FRACIN  measured by \TRACTOR for each object in each of the three bands ($grz$). These are defined and discussed in \S~\ref{subsec:fraccuts}.

\begin{figure*}
	\includegraphics[width=12cm]{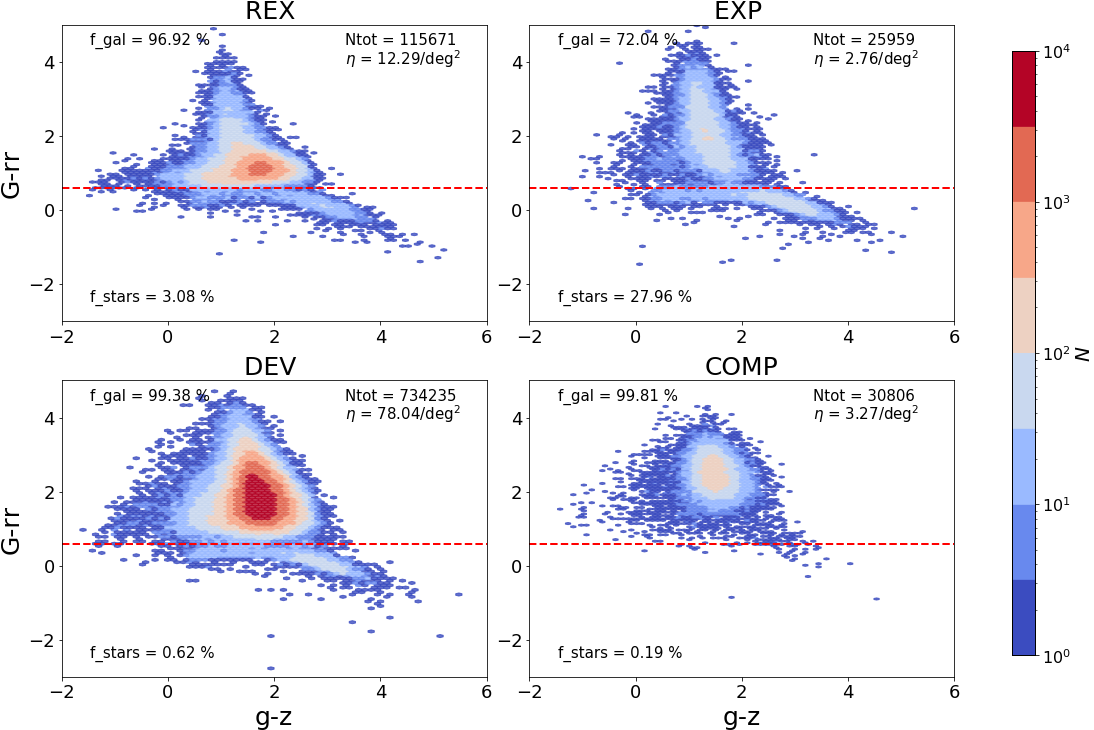}
	\includegraphics[width=5cm]{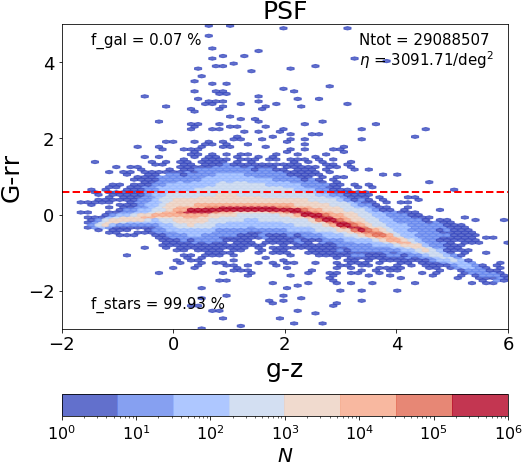}
    \caption{Separately for objects classified by \TRACTOR as type \REX, \EXP, \DEV \COMP and \PSF we show the difference between the \GAIA (PSF) magnitude $G$ and total non-dust corrected r-band model magnitude measured by \TRACTOR, $rr$ versus \TRACTOR extinction corrected $g-z$ colour. All the objects plotted have passed the geometrical and pixel cuts detailed in Fig.~\ref{fig:flow1}, and all but the star-galaxy classification cut of the  photometric-based cuts detailed in Fig.~\ref{fig:flow2}.
    The plots show objects that have been cross-matched between LS \DReight objects and \GAIA DR2. Each panel shows a different morphological class, as labelled,  according to the best-fitting light profile assigned by \TRACTOR. The red-dashed line indicates our adopted division at $G-rr = 0.6$ with stars below and galaxies above the line. The colour in the plots shows the number counts of objects in an hexagonal cell, ranging from $1$ to $10\,000$,  except for the case of \PSF-type objects,  in which case the colour scale covers the range from $1$ to $1$ million as indicated in the colour bars. We display the fraction of galaxies and stars according to this classification at the top-left corner and bottom-left corner respectively. The total number of objects ($N_{\rm tot}$) in each plot and the target density ($\eta$) this represents is displayed in the top-right corner.}
    \label{fig:Grr-gz} 
\end{figure*}

In Fig.~\ref{fig:flow2} we show the second part of the \BGS target selection flow chart. This flow chart focuses on the photometric selection cuts and starts from where the previous flow chart (Fig.~\ref{fig:flow1}), showing the spatial cuts, left off. 
 The \BGS catalogue, in the \DECaLS subregion, ends up having a reduced area of $9\,401$ deg$^{2}$ out of the initial $9\,717$ deg$^{2}$, and target densities of $846$ objects/deg$^{2}$ and $578$  objects/deg$^{2}$ for \BGSB and \BGSF respectively. 

\subsection{Star-galaxy separation}\label{subsec:star_galaxy}

The classification of images as star or galaxies is an old problem that is of great importance when defining target catalogues 
for the efficient use of multi-object spectrographs. Sophisticated techniques are employed which 
 include algorithms using machine learning methods applied to both colour and morphological information e.g. artificial neural networks (\citealt{1992AJ....103..318O}; \citealt{1996A&AS..117..393B}), support vector machines (\citealt{2012ApJ...760...15F}) and decision trees (\citealt{1995AJ....109.2401W}). 
\TRACTOR uses a rigorous statistical approach to determine the best fitting light profile model to each object. In this way it classifies objects as either point sources (\PSF) or extended sources (\DEV, \EXP, \COMP or \REX). However, this pipeline is not infallible and it is inevitable with ground based seeing that some compact galaxies will be misclassified as being of \PSF type rather than extended. As we want to avoid incompleteness that depends on the variable seeing of the images we have instead made use of the space based high angular resolution \GAIA photometry to distinguish point sources from extended sources. This is possible for the \BGS as virtually\footnote{\GAIA DR2 is complete between $12<G$-mag$<17$.} all stars brighter than the \BGS magnitude limit of $r<20$ are bright enough to be detected by \GAIA.




\begin{figure*}
	\includegraphics[width=12cm]{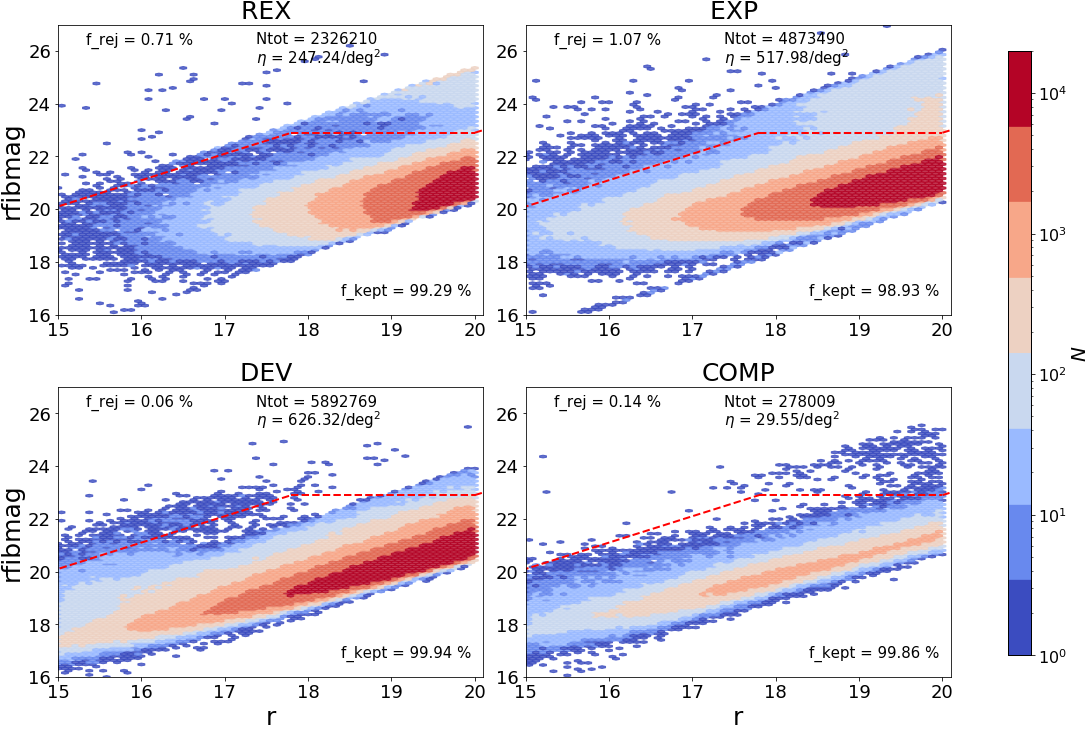}
	\includegraphics[width=5cm]{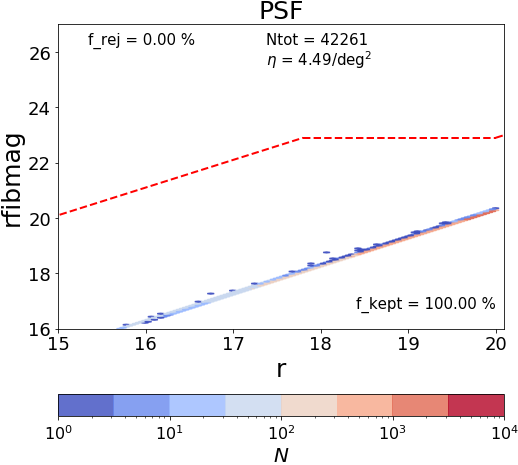}
    \caption{\BGS galaxies in the $r$-band total magnitude (x-axis) versus $r$-band fibre magnitude (y-axis) plane in the \LS \DReight. The results are divided into the five different \TRACTOR best-fitting light profile models, as labelled at the top of each panel. The colour bar shows the number counts of objects in an hexagonal cell covering the range from $1$ to $20\,000$ for four of the light profile models with the exception of \PSF-type galaxies, in which case the scale covers $1$ to $10\,000$. The red-dashed line shows the fibre magnitude cut (\FMC): we reject every object that is above this threshold. The numbers shown in  top-left and bottom-right corners give the fraction of galaxies rejected and kept, respectively, while the number in the top-right corner shows the total number of galaxies ($Ntot$) and the corresponding target density ($\eta$).}
    \label{fig:fmc}
\end{figure*}

The \GAIA DR2 catalogue  \citep{2018A&A...616A...1G} that we use is primarily a catalogue of stars but has some galaxy and quasar contamination as reported by \citet{2019MNRAS.490.5615B}. This means we cannot simply classify all of the \BGS objects that are in \GAIA as stars. However, by comparing \TRACTOR magnitude measurements with the higher spatial resolution magnitude measurements from \GAIA we can determine which objects have extended light profiles. The \GAIA magnitudes are computed assuming all objects are point sources. This results in accurate magnitudes for stars but magnitudes that are systematically fainter than the associated total magnitudes for sources that are extended compared to the $\sim 0.4$~arcsec \PSF achieved by \GAIA. In contrast, the model magnitudes computed by \TRACTOR should capture more fully the total magnitude of the object. Consequently, if \GAIA and \TRACTOR magnitudes were measured in the same band, we would expect them to agree for point sources but for the \TRACTOR magnitude to be brighter than the \GAIA magnitude for extended sources. We would even expect this to be true for extended objects that \TRACTOR mis-classifies as PSF since the wide, ground-based PSF of \TRACTOR would capture more of the total flux than the narrow PSF of \GAIA. The complication is that the \GAIA $G$ band is a much wider filter than the \DESI $r$ band, but as we shall see, the colour dependence is weak.


Based on these considerations we define \TRACTOR objects with $r<20$ as being galaxies if either of the following two conditions is met:
\begin{itemize}
    \item The object is not in the \GAIA  catalogue.
    \item The object is in the \GAIA catalogue but has $G-rr~>~0.6$.
\end{itemize}
In the above, the $G$-band is the $G$ photometric \GAIA magnitude and $rr$ is the raw $r$-band magnitude from the \LS \DReight\ {\it without} applying a correction for Galactic extinction. This choice is made because the \GAIA magnitude is not corrected for Galactic extinction. The discussion above explains that $G$ and $rr$ magnitudes are measured in different effective apertures and so the quantity $G-rr$ should be thought of as a measure of how spatially extended an object is and not its colour.  The first criterion above is satisfied by most (93 per cent) of the \BGS objects. It leaves very little stellar contamination in the  \BGS, as essentially any star brighter than $r=20$ is bright enough to be detected and catalogued by \GAIA.  The second criterion is required to keep the \BGS completeness high by not rejecting galaxies that are in the \GAIA catalogue.

In Fig.~\ref{fig:Grr-gz} we show the $G-rr$ versus $g-z$ plane for objects in \GAIA DR2 that are matched with objects in the \LS \DReight. The panels show different objects as classified by the \TRACTOR model fits (i.e., \PSF, \COMP, \DEV, \EXP, \REX). The cross-matched objects have been subject to all the \BGS cuts (i.e. both spatial and photometric) with the exception of the star-galaxy separation itself. For objects classified by \TRACTOR as \PSF-type, we can see the stellar locus around $G-rr=0$ with a weak colour dependence. For the extended sources (i.e., \COMP, \DEV, \EXP, \REX), we see part of the galaxy locus\footnote{We have to remember that Fig.~\ref{fig:Grr-gz} only includes stars and galaxies that are cross-matched between \LS \DReight and \GAIA DR2.} in the upper part of the plot, just above $G-rr=0$. 

From Fig.~\ref{fig:Grr-gz} we can see that the assignment of the best fitting \TRACTOR model supports our \GAIA classification using $G-rr>0.6$, but we can still see some remnants of the stellar locus for objects that have not been assigned PSF-type  by \TRACTOR. For the objects classified PSF-type by \TRACTOR we see in the right-most panel of Fig.~\ref{fig:Grr-gz} that $99.93$ per cent fall on the stellar side of our $G-rr$ cut.  For the objects classified by \TRACTOR as the extended types (\REX, \DEV and \COMP) the stellar contamination (i.e. objects with $G-rr<0.6$) is at most $3.1$ per cent. However, the contamination of the \EXP-type objects is approaching $30$ per cent. 

The \BGS target selection has the expected surface density after applying the star-galaxy separation. From the spatial cut flow chart in Fig.~\ref{fig:flow2}, we find a bright target density of $868.91$ objects/deg$^{2}$ and a faint target density of $598.82$ objects/deg$^{2}$. Rejected \GAIA stars have a target density of $2\,804.01$ objects/deg$^{2}$ bright stars and $622.80$ objects/deg$^{2}$ faint stars.

\subsection{Fibre magnitude cut}\label{subsec:fibrecut}

In order to reduce the number of image artefacts and fragments of `shredded' galaxies that would otherwise be classified as \BGS targets we apply a cut on the fibre magnitude that is defined as a function of $r$-band magnitude as follows:
\begin{equation}\label{eq:fmc}
{\rm rfibmag} <
   \begin{cases}
     22.9 +  (r-17.8)  & \text{for } r < 17.8 \\
     22.9                          & \text{for } 17.8 < r < 20 
   \end{cases}
\end{equation}
where ${\rm rfibmag}$ is the magnitude of the predicted $r$-band fibre flux and $r$ is the total $r$-band magnitude, both extinction corrected. The location of this cut was guided by inspecting postage stamp images of a selection of the objects with the faintest fibre magnitudes with the aim of rejecting objects that appear to be artefacts while retaining nearly all of the genuine galaxies. In addition, at the bright end our threshold was guided by the location of spectroscopically confirmed \GAMA galaxies, as discussed further in Section~\ref{subsec:gama_validation}. 
Fig.~\ref{fig:fmc} shows the distribution of the \BGS objects in the  ${\rm rfibmag}$ vs.  ${\rm rmag}$ plane, with a separate panel for the different \TRACTOR classes, and a red-dashed line indicating the location of the {\it fibre magnitude cut} (hereafter \FMC). 
In the first four panels we can see that the galaxy locus has a tight core and, in general, is well below the \FMC. The \FMC removes 1.2 per cent of the objects classified as \EXP and even smaller fractions of the other light profile classes.  

All \BGS objects in the \PSF class lie on a stellar locus. Whether all these objects are stars or whether this is an artefact of \TRACTOR only fitting the \PSF model to \GAIA sources with low astrometric excess noise (AEN) is revisited in Section~\ref{subsec:gama_validation}, where we compare our classification with that of the \GAMA DR4 survey. The stellar locus is also visible in the other photometric classes indicating there is some stellar contamination in our sample, but it is at a very low level.


In summary the adopted \FMC rejects a further $23.17$ objects/deg$^{2}$ of which $11.72$ are in \BGSB and $11.45$ are in \BGSF from the objects that have passed the previous cuts which include the rejection of stars by our star-galaxy classifier.

\subsection{Colour cuts}\label{subsec:colourcuts}

\begin{figure}
	\includegraphics[width=\columnwidth]{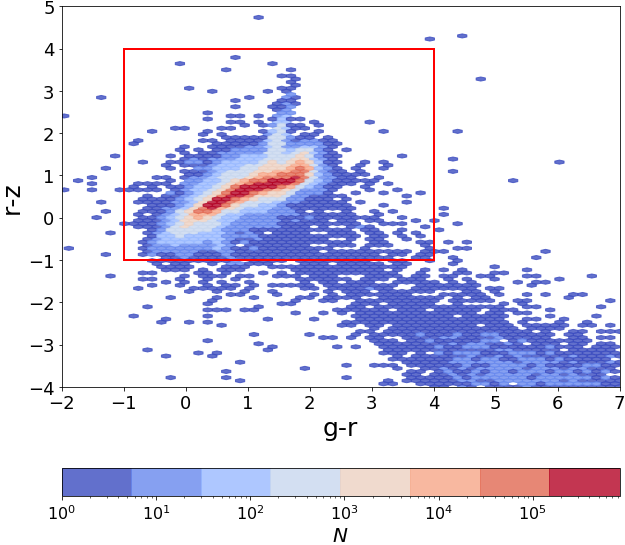}
    \caption{Colour-colour distribution showing $g-r$ vs. $r-z$ for \BGS objects without applying the \CC. The colour bar shows the number counts of objects in an hexagonal cell covering the range from $1$ to $800\,000$. The solid red box shows \CC defined in Equation~\ref{eq:colcut}. Sources outside of this box are excluded from the \BGS.}
    \label{fig:cc}
\end{figure}

An efficient way of rejecting further spurious targets from the \BGS is to reject objects with bizarre colours. 
The limits we impose to reject outliers are:
\begin{eqnarray}
    -1 &< \,g-r &< 4 \nonumber \\
    -1 &< \,r-z &< 4.
    \label{eq:colcut}
\end{eqnarray}
Fig.~\ref{fig:cc} shows the $g-r$ vs. $r-z$ colour-colour distribution of the objects retained in BGS if all but the colour cut (\CC) were applied. The red box indicates the colour range we keep.
We can see from this plot that the locus of normal galaxy colours lies well within the range we retain and the cuts are only removing objects/artefacts with bizarre colours.  It is evident that some stellar contamination remains as the stellar locus can be seen as a spur of objects with very red $r-z$ colours. However the density of objects in this spur, and its blueward extension which overlaps the galaxy locus, is no more than a few objects/deg$^2$ as we shall see in Section~\ref{subsec:gama_validation}.  The 
colour cuts (\CC) we apply reject an additional $6.7$  objects/deg$^{2}$, with $2.66$ in \BGSB and $4.04$ in \BGSF.

\subsection{Quality cuts}\label{subsec:fraccuts}
\label{sec:QC}

Each object in the \TRACTOR catalogue has three measures of the quality of its photometry recorded in each of the three bands ($grz$). These are:
\begin{itemize}
    \item FRACKMASK (FM): The profile-weighted fraction of pixels masked in all observations of the object in a particular band. This quantity lies in the range $[0,1]$. High values indicate that most of the flux of the fitted model lies in pixels for which there is no data due to masking and so the measurement is unreliable. 
    \item FRACIN (FI): The fraction of the model flux that lies within the set of contiguous pixels (termed a `blob') to which the model was fitted. FRACIN is close to unity for most real sources. Low values indicate that most of the model flux is an extrapolation of the model into regions in which no data was available to constrain it.  
    \item FRACFLUX (FF): The profile-weighted fraction of the flux from other sources divided by the total flux of the object in question. FRACFLUX is zero for isolated objects but can become large for faint objects detected in the wings of brighter objects that are nearby. 
\end{itemize}

Once the other cuts have been applied, in particular, the cut on \NOBS and the \BS mask, the distribution of each of these quantities is tightly peaked around the favoured values of \FRACMASKED $\approx 0$, \FRACIN $\approx 1$ and
\FRACFLUX $\approx 0$. However, each quantity has a distribution with a fairly featureless tail that extends out to less desirable values. There are also clear correlations between the three quantities for a given photometric band and in some cases between photometric bands.  The choice of the best set of thresholds to reject outliers is not trivial. We have adopted the following quality cuts (\QCs):
\begin{eqnarray}
    \rm{\FRACMASKED}\_i &&< 0.4, \nonumber \\
    \rm{\FRACIN}\_i  &&> 0.3, \nonumber \\
    \rm{\FRACFLUX}\_i  &&< 5, \hspace{1cm} {\rm where} \, i = \textrm{$g$, $r$ or $z$},
\label{eq:qualcut}
\end{eqnarray}
based on visual inspection of postage stamp images.


As mentioned in  Section~\ref{subsec:pix_masking}, we find that the objects flagged by the \TRACTOR quantity \ALLMASK are essentially a subset of the objects that are rejected by applying the quality cuts listed in Eqn.~\ref{eq:qualcut}. While cutting on \ALLMASK would have the advantage that it could also be applied to the randoms, we find that it is important to apply the \QCs to remove spurious objects that are missed by the other cuts. For instance, some spurious objects that are outliers in either the fibermag vs. mag plane or in the colour-colour space that just pass the \FMC and \CC are removed by considering \FRACMASKED or \FRACIN. 

As shown in the flow chart, Fig.~\ref{fig:flow2}, the \QCs reject an additional $14.11$ objects/deg$^{2}$ of which $\sim 60$ per cent are removed by \FRACFLUX, $\sim 45$ per cent by \FRACMASKED and $\sim 7$ per cent due to \FRACIN. The overlap between the \FRACMASKED, \FRACIN and \FRACFLUX cuts is minimal, with only $1.05$ objects/deg$^2$ for objects with $r < 19.5$, and in round $0.15$ objects/deg$^2$ for objects with $19.5 < r < 20$ being rejected by more than one of the cuts.
Separately for \BGSB and \BGSF, we show the target density of objects rejected by these cuts after applying all the previous cuts. The largest overlap between these cuts is between
 \FRACMASKED and \FRACFLUX for \BGSB, but even here it amounts to less than $1$ object/deg$^2$.
 For \BGSF this overlap is small,  $0.11$ object/deg$^2$, and there is no overlap with \FRACIN.

\medskip

In Appendix~\ref{app:galview} we present another version of the selection cut flow chart in which the cuts are applied in a different order. There we give a galaxy view of the target selection by first applying the star-galaxy classification so that all the subsequent cuts apply only to galaxies. The final selected sample
which comprises of $845.5$~galaxies/deg$^2$ in \BGSB and $577.9$~galaxies/deg$^2$ in \BGSF, 
is exactly the same, as the order of the cuts does not matter. The objects rejected by each filter, however, does change as many objects are rejected by more than one filter.  To illustrate this point we have also  swapped the order of the \FMC and \QCs cuts so one can see how these influence one another.

\begin{figure*}
	\includegraphics[width=17cm]{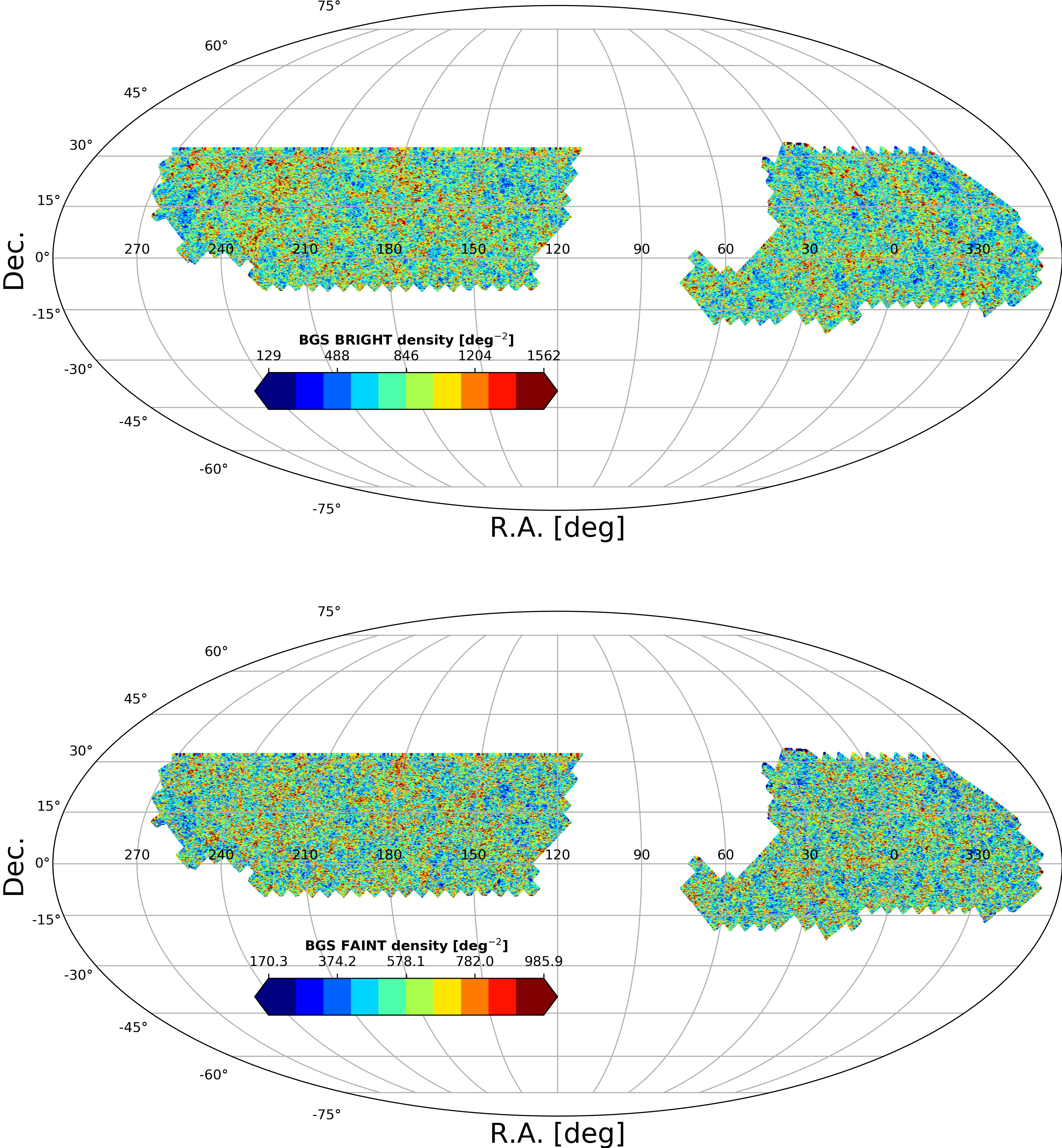}
    \caption{The distribution on the sky of the \BGSB (upper map) and \BGSF (bottom map) target density in objects/deg$^2$, computed on a HEALPix grid with a resolution of $N_{\rm side} = 256$. The mean densities are $846$ and $579$ objects/deg$^2$ for the bright and faint \BGS respectively.}
    \label{fig:skymap_densities}
\end{figure*}

\section{Catalogue properties}
\label{sec:cat_properties}

\begin{table}
\caption{The \BGS target densities for each of the \TRACTOR best-fitting photometric models. 
The first column labels the photometric model. The next three columns list the surface density of objects per deg$^{2}$ for the \BGSB and \BGSF samples separately and their combined sum. The area covered by the DECaLS portion of the \BGS is $9,401$ deg$^{2}$. }
\centering
\begin{tabular}{l r r r} 
 \hline
 {\bf Model} & $\eta_{\rm bright}$  & $\eta_{\rm faint}$  & $\eta_{\rm overall}$  \\
 &   [deg$^{-2}$] &    [deg$^{-2}$] & [deg$^{-2}$] \\
 \hline
 \DEV  & $427$ & $202$ & $629$ \\
 \EXP  & $284$ & $230$ & $514$ \\
 \REX  & $104$ & $141$ & $246$ \\
 \COMP  & $27$ 	& $3$ &	 $31$ \\
 \PSF  & $3$ &	$2$ & $5$\\
 \hline
 Total & $846$ & $578$ & $1423$ \\
 \hline
\end{tabular}
\label{tab:bgs_outline}
\end{table}

The final \BGS catalogue in the \DECam region in the South Galactic Cap (\SGC) covers the declination range $-17 \lsim {\rm DEC} \lsim 32$~degrees, and in the North Galactic Cap (\NGC) the range $-10\lsim {\rm DEC} \lsim 32$~degrees. The \BGS has a total of $13,378,062$ galaxies of which $7,944,975$ are in \BGSB and $5,433,087$ are in \BGSF. The total area covered by the \BGS in the \DECaLS subregion defind by the footprint of the tiles in Fig.\ref{fig:decals_in_desi} and after accounting for the spatial cuts described in Section~\ref{sec:spatial_masking} is $9\,401$ deg$^{2}$. In Table~\ref{tab:bgs_outline} we list the target density of the \BGS catalogue for each of the best-fitting photometric models used in \TRACTOR.

In Fig.~\ref{fig:skymap_densities} we show the  \BGSB and \BGSF sky map densities computed with the HEALPix scheme using
\begin{eqnarray}
    \eta_{i} &=& N^{\rm BGS}_{i}/A_{\rm eff} , \label{eq:eta_sys}\\ \nonumber
    A_{\rm eff} &=& N^{R}_{i}/\eta^{R} , 
\end{eqnarray}
where for each pixel $N^{\rm BGS}_{\rm i}$ is the number of \BGS targets, $A_{\rm eff}$ is the effective area computed from the number of randoms, $N^{\rm R}_{\rm i}$, and the total surface density of the randoms, without any masking, is $\eta^{R}=15,000$ objects/deg$^2$. We use a HEALPix grid of $N_{\rm side} = 256$ giving a pixel area of $A_{\rm pix}=0.052$ deg$^2$. The appearance of the density fluctuations is very similar in the two disjoint regions and show no variation with galactic latitude. We look more closely at systematic variations in the target density in Section~\ref{subsec:systematics}.

\subsection{Cross-comparison with GAMA}\label{subsec:gama_validation}

The main target sample in GAMA \citep{10.1093/mnras/stx3042} is a complete sample of galaxies with \SDSS Petrosian $r$-band magnitude brighter than $r=19.8$. The Petrosian magnitude is measured within a circular aperture of twice the Petrosian radius, where the radius is computed using the $r$-band surface brightness profile \citep{2008ApJS..175..297A}.  The \GAMA photometric selection is very similar to that of DESI \BGS and so we expect a very similar redshift distribution as GAMA which has median of $z=0.2$ and a 90 percentile value of  $z =0.5$.


Star-galaxy separation in \GAMA was conservative in that it aimed for very high completeness at the expense of some stellar contamination.  These properties combined with its very high spectroscopic completeness 
(high quality redshift have been obtained for more $98.85$ per cent of the \GAMA targets) make it a nearly ideal "truth table" from which to assess the completeness of the BGS target selection and measure the expected redshift distribution of the \BGSB sample.
Below we make use of GAMA to examine various aspects of our \BGS catalogue. In Sec.~\ref{subsubsec:GAMAmagz} we
compare the $r$-band phototometry of the matched objects and determine the redshift distribution of the \BGS galaxies that match with galaxies in the \GAMA survey.  
Section~\ref{subsubsec:psf_bgs} explores an issue related to \TRACTOR only providing PSF photometry for some of the BGS galaxies.  In Section~\ref{subsubsec:assess_bgs} we assess incompleteness in \BGS relative to \GAMA and quantify how much is caused by each of the various geometric and photmetric selections.

\subsubsection{Magnitude definition and redshift distribution}\label{subsubsec:GAMAmagz}

We match the \GAMA Main Survey DR4 galaxy catalogue \citep{2012yCat..74130971D, 2015MNRAS.452.2087L, 10.1093/mnras/stx3042}, which is defined by a Petrosian magnitude (\RPETRO) limit of $r=19.8$, to the \BGS target catalogue. We use a maximum linking length of $1$ arcsec to match them. The mean separation of the matches we find 
is $0.093$ arcsec with a $1\sigma$ dispersion of $0.091$~arcsec. We focus on three of the five \GAMA fields: G09, G12, G15. We omit G02 as this \GAMA field is only partially within the \DECaLS footprint, and G23 is far to the south. The redshift completeness of the main \GAMA survey is extremely high  in the sense that $98.85$ per cent of the objects in the catalogue yield redshifts with a quality flag  \NQ$\geq 3$.

The \GAMA spectroscopic redshifts can be used to reliably reject stars with a cut at $z=0.002$. In what follows we restrict our \GAMA catalogue to the spectroscopically confirmed galaxies ($\sim98$ per cent of the full catalogue).
The area of each of the \GAMA fields considered is $59.98$~deg$^2$ which means that our matched sample has a total area of $\sim180$ deg$^2$. The overall density of sources that are cross-matched between \BGS and \GAMA galaxies is $\sim970$ objects/deg$^{2}$ with a mean redshift of $z=0.224$.

\begin{figure*}
	\includegraphics[width=17cm]{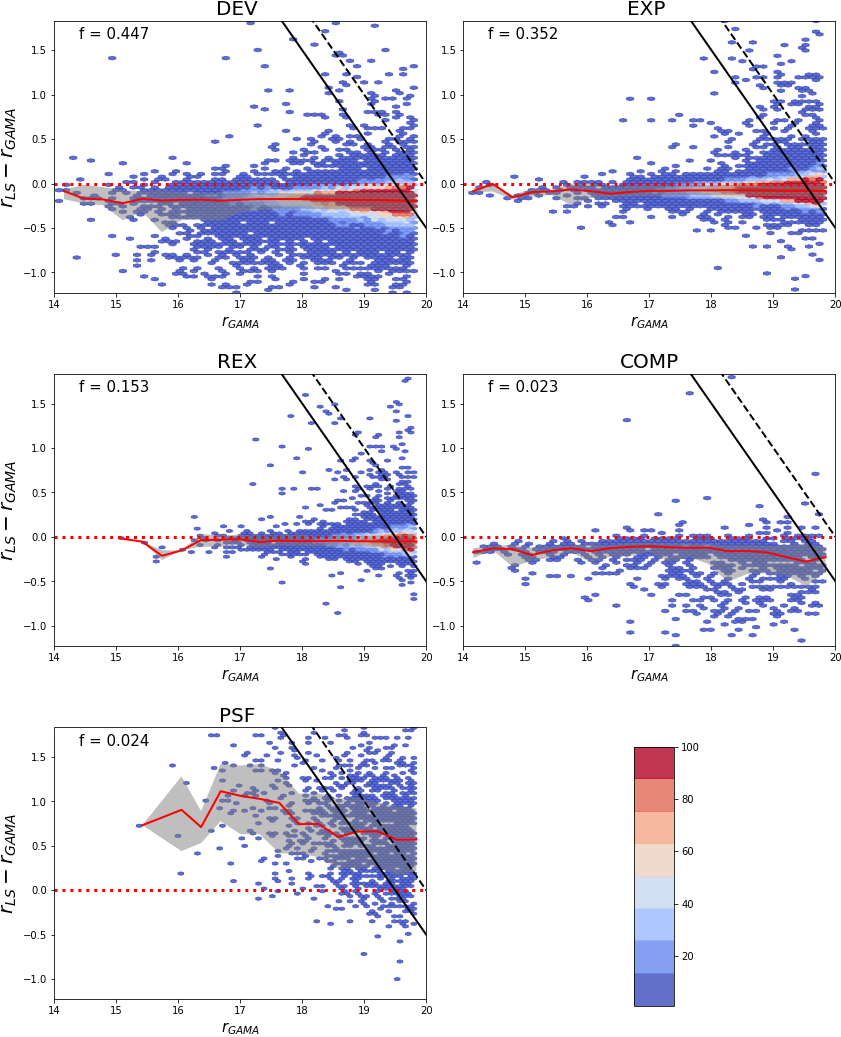}
    \caption{The $r$-band total magnitude in the \LS ($r_{\rm LS}$) vs the SDSS $r$-band Petrosian magnitude in \GAMA ($r_{\rm GAMA}$) for \LS \DReight objects cross-matched with \GAMA. Each plot corresponds to one of the five photometric model fits assigned by \TRACTOR. The red solid line shows the median value of $r_{\rm LS}-r_{\rm GAMA}$ as function $r_{\rm LS}$; the gray shading shows the $20$ to $80$ percentile range; the dashed black line shows the limiting magnitude of $r_{\rm LS} = 20$ for \BGS and the solid black line shows limiting magnitude of $r_{\rm LS} = 19.5$ for \BGS. The colour bar shows the number counts of objects in an hexagonal cell covering the range from . The fraction of \LS \DReight objects plotted out of the total number matched with \GAMA is shown in the top-left corner of each panel.}
    \label{fig:bgs_gama_magdiff}
\end{figure*}


For this matched catalogue, Fig.~\ref{fig:bgs_gama_magdiff} compares the \DReight
$r$-band total magnitude ($r_{\rm \LS}$) with the  Petrosian $r$-band magnitude from \GAMA ($r_{\rm GAMA}$) by plotting $r_{\rm \LS} - r_{\rm GAMA}$ vs $r_{\rm GAMA}$. To see how this difference depends on galaxy morphology, we divide the \LS galaxies into the five photometric classes assigned by \TRACTOR. In each panel we show the fraction of matched galaxies in each \TRACTOR model fit class; \DEV and \EXP classes together
make up 80 per cent of the sample and the \PSF class just 2.5 per cent.
We mark on the plot the $r_{\rm \LS}<20$ limit of \BGS, but note this has not been applied when defining the \LS sample that was matched to \GAMA.

\begin{figure}
	\includegraphics[width=\columnwidth]{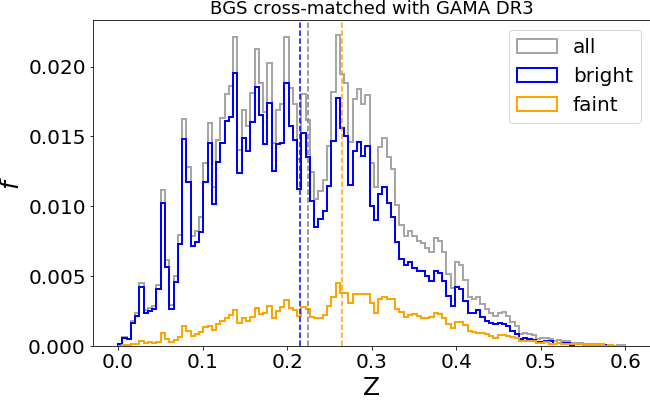}
    \caption{The redshift distribution of \BGS objects cross-matched with \GAMA DR4 broken into bright ($r < 19.5$, blue) and faint ($19.5 < r < 20$, orange) galaxies according to the \BGS $r$-band. The gray histogram shows the overall redshift distribution of \BGS galaxies cross-matched with \GAMA. The mean redshift values for each distribution are: $0.215$ for the  bright sample (dashed blue line), $0.265$ for the faint sample (dashed orange) and $0.224$ for all galaxies (dashed gray).}
    \label{fig:bgs_gama_z}
\end{figure}

\begin{figure*}
	\includegraphics[width=17cm]{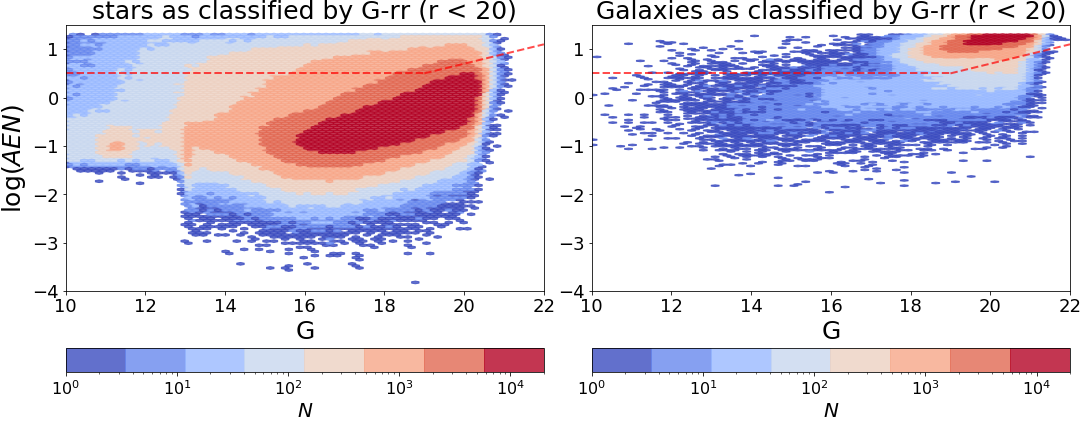}
    \caption{The \GAIA Astrometric Excess Noise parameter (AEN) versus $G$-band magnitude.
     The top panel shows \GAIA objects classified as stars by \BGS and the bottom those classified as galaxies.
    Both plots only show \GAIA objects with magnitudes limit of $r < 20$. The red dashed-line represents the threshold limit for the AEN classification used in TRACTOR, therefore everything below the line is a star and everything above is a galaxy according to the AEN classification. The colour bar shows the number counts of objects in an hexagonal cell covering the range from $1$ to $20\,000$.}
    \label{fig:G_logAEN}
\end{figure*}

Differences in the effective passbands of the $r$-band filters of the \LS and \SDSS result in offsets in $r_{\rm \LS} - r_{\rm GAMA}$ of around $-0.05$ and $-0.1$ for blue and red galaxies respectively \citep{Dey:2019}. One also has to consider the difference in magnitude definitions which contributes the more to this magnitude offset. 
To the extent that the best fit profiles accurately describe the actual light profiles of the objects,  \LS provides total magnitudes. In contrast, the \SDSS Petrosian magnitudes used by \GAMA 
quantify only the flux within twice the Petrosian radius \citep{2001AJ....121.2358B}. The fraction of the flux within this aperture depends on the light profile. For EXP profile it captures 99.4 per cent, but for the DEV profile which, is more sharply peaked but with broader wings, only
82 per cent is captured. It is these differences in definition which largely drive the differences in median offsets we see in the DEV, EXP, REX and COMP classes. In all these cases the LS magnitude is brighter (more negative) than the GAMA magnitude with median offsets being
$-0.085$~magnitudes for EXP  and $-0.188$~magnitudes for DEV.
In contrast for the
\PSF case the median $r_{\rm \LS}-r_{\rm \GAMA}$ is positive, which means that the \LS PSF model magnitude captures less flux than the \GAMA Petrosian magnitude. For true point sources we would expect these two magnitudes to be almost equal. The positive difference appears to happen because \TRACTOR force fits \PSF models to sources that are actually extended (deemed extended by our \GAIA based star-galaxy separation) and consequently underestimates their fluxes. The reason this happens is discussed in Section~\ref{subsubsec:psf_bgs}.

If we take account of the scatter between the \BGS and \GAMA magnitudes we can use \GAMA to assess the level of contamination in the \BGS catalogue.  If we treat \GAMA as being a 100 per cent complete galaxy catalogue then any objects in \BGS that are not in \GAMA would be contamination in the form of stars or image artefacts.  This is not true at $r=20$ as here some \BGS objects will not be in \GAMA simply because of the $r_{\rm petro}<19.8$ magnitude limit in \GAMA. This can be seen in Fig.~\ref{fig:bgs_gama_magdiff} from the location of the 
$r_{\rm LS}=20$ dashed line relative to where the \GAMA data truncates at $r_{\rm GAMA}=19.8$.
To avoid this problem if we apply a brighter magnitude limit $r<r_{\rm lim}$ to \BGS then for a broad range of $18.5\lesssim r_{\rm lim} \lesssim 19.3$ we find that $\sim 3$ percent of \BGS objects are not matched with \GAMA galaxies. This sets an upper limit (in this magnitude range) of 3 per cent contamination in \BGS as \GAMA itself may not be 100 per cent complete.

Fig.~\ref{fig:bgs_gama_z} shows the distribution of redshifts for \BGS objects that have been cross-matched with \GAMA galaxies. The overall distribution is shown along with those for the  \BGSF and \BGSB.
We expect this distribution to be representative of the \BGSB sample as we can see from 
Fig.~\ref{fig:bgs_gama_magdiff} that incompleteness caused by the GAMA magnitude limit to be very small. However the redshift distribution plotted for \BGSF is more strongly affected by the GAMA magnitude limit and its true redshift distribution is expected to be more extended.

\subsubsection{Galaxies with TRACTOR type PSF}\label{subsubsec:psf_bgs}

To avoid stars being classified as extended sources \TRACTOR uses a catalogue of stars from \GAIA to pre-select a set of objects on which it will only allow \PSF fits. 
The \GAIA objects for which it does this are based on the following cut on the 
\GAIA astrometric excess noise parameter , AEN,
\begin{eqnarray}
      {\rm AEN} < 10^{0.5}, & G \leq 19\\ \nonumber
      {\rm AEN} < 10^{0.5 + 0.2(G - 19)}, & G \geq 19,
\end{eqnarray}
where $G$ is the \GAIA photometric $G$-band.
The AEN can be used as measure of whether a source is extended as for extended sources the astrometric measurements are noisier than one would expect for a point source.

 In contrast, in the \BGS we use the difference between the \GAIA $G$-band magnitude and the \TRACTOR raw $r$-band magnitude, $rr$, (not corrected for extinction) as a measure of how extended the object is (see Section~\ref{subsec:star_galaxy}). 
In Fig.~\ref{fig:G_logAEN} we have plotted $\log({\rm AEN})$ versus $G$  separately for objects classified as stars and galaxies by our $G-rr$ classifier.  The threshold adopted by \TRACTOR can be seen to separate the bulk galaxies from the stars. For $96$ objects/deg$^2$ the two criteria agree the object is a galaxy,
but the distributions are extended and the agreement is not perfect. 
There are
$36$ objects/deg$^2$ that the AEN criterion classifies as galaxies which $G-rr$ classifies as stars. More problematic are the  $5$ objects/deg$^2$ that the AEN criterion classifies as stars which $G-rr$ classifies as galaxies.
This is an issue as it means some objects that are classified as galaxies in the \BGS are treated by  \TRACTOR as stars and only have a \PSF light profile fitted.  Overall in the \BGS there are $5$~objects/deg$^{2}$ with \PSF type within the \DECaLS footprint (see Table~\ref{tab:bgs_outline}). These objects have fibre magnitudes that are consistent with the locus of stars in Fig.~\ref{fig:fmc} which makes us question if they really are galaxies.  
We investigate this below by making use of \GAMA to determine whether or not they are galaxies.

\begin{table}
\caption{The surface density of  \PSF-type objects in the \BGS in the G09, G12 and G15 \GAMA fields combined before ($\eta_{\rm BM}$) and after ($\eta_{\rm AM}$) cross-matching with \GAMA (top half of table). The bottom half of the table shows the surface density and percentage of objects in disjoint subsamples of the \PSF-type \BGS sample, as listed in the first column: objects that are not in \GAIA, objects that the AEN scheme classifies as stars and those that the AEN scheme classifies as galaxies.}
\centering
\begin{tabular}{lllll}
 \hline
 {\bf Sample} & $\eta_{\rm BM}$ &   & $\eta_{\rm AM}$ &   \\
       &   [deg$^{-2}$]  &    &   [deg$^{-2}$]  &  \\
       \hline
 \PSF-type BGS & $4.10$ & & $1.76$ &  \\
 \hline
 {\bf Subsample} & $\eta_{\rm BM}$  & \%$_{\rm BM}$ & $\eta_{\rm AM}$ &  \%$_{\rm AM}$ \\
      &  [deg$^{-2}$]   &     & [deg$^{-2}$]    &    \\
 \hline
 Not in \GAIA  & $1.72$ & $42.0$ & $0.04$ & $~~2.3$  \\
 \GAIA AEN star  & $2.26$ & $55.2$  &  $1.69$ & $96.4$   \\
 \GAIA AEN galaxy & $0.11$ & $~~2.8$  & $0.02$ & $~~1.3$   \\
 \hline
\end{tabular}
\label{tab:bgs_psf_gama}
\end{table}

\begin{figure}
	\includegraphics[width=\columnwidth]{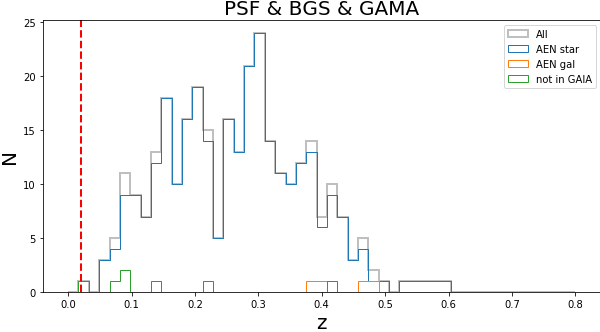}
    \caption{Redshift distribution of \PSF-type \BGS galaxies cross-matched with galaxies from three \GAMA fields (G09, G12, G15). Redshifts are taken from \GAMA DR4. The four distributions correspond to the matched sample (gray) and the disjoint subsamples comprising galaxies not in \GAIA (green), and stars (blue) and galaxies (red), as defined by the AEN classification. The red dashed line marks the redshift $z=0.002$; objects with redshifts smaller than this are stars.}
    \label{fig:matched_sample_z}
\end{figure}

First, we restrict our attention to the  $180$ deg$^2$ of our matched \GAMA catalogue. 
The \BGS \PSF-type galaxies (main sample) have a density of $4.10$ objects/deg$^{2}$, somewhat less than the  $5$ objects/deg$^{2}$ which is the average over the full \DECaLS area. This reduces further to $1.76$ objects/deg$^{2}$ after cross-matching with \GAMA. 
We further subdivide these two cases (\BGS PSF type and \BGS \PSF type cross-matched with \GAMA) into three disjoint sub samples: i) those that are not in \GAIA, ii) those that are in \GAIA and which are classified using the AEN value as stars, and iii) 
those that are in \GAIA and which are classified using the AEN value as galaxies.

The subsample sizes are reported in Table~\ref{tab:bgs_psf_gama}, where we give the surface density of objects before and after the cross-match with GAMA ($\eta_{\rm BM}$ and $\eta_{\rm AM}$) along with the percentage of the total number of objects represented by each subsample. This shows that $\sim 96$ per cent of the \BGS \PSF-type cross-matched with GAMA are \GAIA AEN stars, which represents the $\sim 55$ per cent in the non-matched sample. For the remained $45$ per cent in the non-matched sample, \GAMA is not reliable to assess this as only $3.6$ per cent of those are cross-matched with \GAMA. Fig.~\ref{fig:matched_sample_z} shows the \GAMA redshift distribution for the \BGS \PSF-type cross-matched with \GAMA broken into the three clases shown in Table~\ref{tab:bgs_psf_gama}. These objects shown a redshift distribution very similar to that of the full BGS sample. The reason for this mis-classification lies in the fact that for objects classified by the \GAIA AEN criterion 
as stars \TRACTOR only fits PSF models.  For the galaxies that this \GAIA AEN criterion falsely classifies as stars \TRACTOR underestimates the total flux of the galaxy resulting in the offset with the \GAMA photometry we saw in the \PSF panel of Fig.~\ref{fig:bgs_gama_magdiff} and putting these galaxies close to the stellar locus in Fig.~\ref{fig:fmc}.


\subsubsection{Incompleteness of \BGS relative to \GAMA}\label{subsubsec:assess_bgs}

\begin{figure}
	\includegraphics[width=\columnwidth]{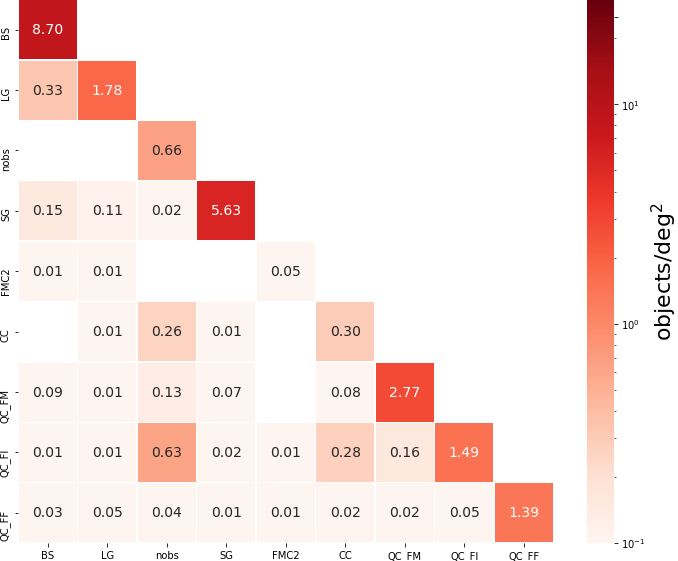}
    \caption{Heatmap showing the target density of \GAMA galaxies ($z > 0.002$) that are missed in the \BGS. The diagonal shows the number of objects per square degree removed by each of the individual spatial and photometric cuts applied in the \BGS while the off-diagonal entries show the densities of objects removed by both cuts labelled on the $x$ and $y$ axes.}
    \label{fig:heatmaps}
\end{figure}

To the depth of GAMA we can assess the completeness of the \BGS catalogue by cross-matching the full depth LS \DReight catalogue with GAMA DR4. 
This cross-match yields a catalogue of 
 $1011$ objects/deg$^{2}$ which represents 
of $99.6 \, \textrm{per cent}$ of the \GAMA catalogue. 
Visual inspection reveals some of the remaining 0.4 per cent are deblending issues where GAMA fragments a galaxy into two objects while TRACTOR keeps it as a single object.
Of the matched objects 
$970$  objects/deg$^{2}$ are in BGS while the other $41$  objects/deg$^{2}$ are excluded from the BGS catalogue by one or other of our selection cuts.

Due to the scatter between \SDSS $r$-band Petrosian magnitude used by GAMA and the \TRACTOR model magnitude used by \BGS (see Fig.~\ref{fig:bgs_gama_magdiff}), the
\BGS $r_{\rm \LS}=20$ magnitude limit excludes $20$ faint GAMA galaxies per square degree.  This leaves $20.8$ objects/deg$^{2}$ in
GAMA that are missing from the \BGS. Whether this represents potential problematic incompleteness in \BGS or just a difference in sample definition depends on which selection cuts remove the objects.
We quantify and discuss this below.

The diagonal elements in Fig.~\ref{fig:heatmaps} indicate the number density of spectroscopically confirmed GAMA galaxies missing from the \BGS catalogue as result of each of the following spatial and photometric cuts: the bright star mask (\BS); the large galaxy mask (\LG); the number of observations (\NOBS); star-galaxy classification (\SG); fibre magnitude cut (\FMC); colour cut (\CC); the \FRACMASKED quality cut (\QCs FM); the \FRACIN quality cut (\QCs FI); the \FRACFLUX quality cut (\QCs FF).
 The off-diagonal entries in Fig.~\ref{fig:heatmaps} show the surface density of \GAMA galaxies that are removed by both of the two cuts indicated by the labels on the $x$ and $y$ axes.

The objects removed by the spatial \BS and \NOBS cuts are benign in that they do not affect BGS clustering measurements. These spatial masks are uncorrelated with BGS galaxy positions and so can be fully accounted for in clustering analyses by applying the same masks to the random catalogue.  The values given in Fig.~\ref{fig:heatmaps} show that these two masks have no overlap and together remove 9.36~objects/deg$^{2}$. 

Applying these two spatial cuts leaves us with 11.43~galaxies/deg$^{2}$ that are in \GAMA but are missed by \BGS. The cuts that remove these objects are almost completely independent.  $5.36$~objects/deg$^{2}$ are removed by the our \SG classification. 
 These objects are close to the cut imposed for the \GAIA star-galaxy separation ($G-rr = 0.6$), but fall on the stellar side. We find that $98 \, \textrm{per cent}$ of these missed \GAMA galaxies are classified as stars according to the \GAIA AEN condition, which means that their photometry has been compromised as \TRACTOR only fitted PSF models.
 If these are extended objects, then their flux as reported by \TRACTOR is a fraction of what it should be and hence their $rr$-magnitude is shifted to fainter values. This results in \BGS galaxies shifting to lower values of $G-rr$, moving them out of  the galaxy locus and into the stellar one. 
If this were
fixed we would expect the residual incompleteness to be 6.07~galaxies/deg$^{2}$, equivalent to 6.07/970 = 0.62 per cent.  The proportions of this produced by the \LG 
\QCs FM, \QCs FI and \QCs FF cuts are
23.5, 41.2, 13.8 and 21.4
per cent respectively with a negligible fraction removed by the \CC and \FMC.


\begin{figure}
	\includegraphics[width=\columnwidth]{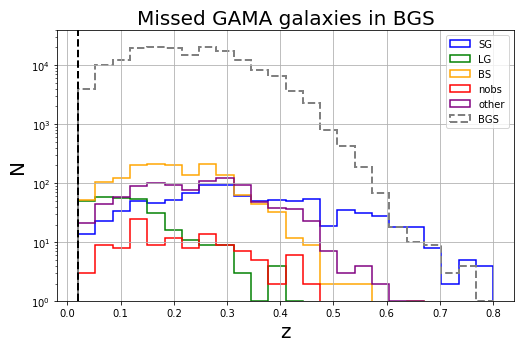}
    \caption{Redshift distribution of the \GAMA galaxies that are not included in the \BGS, with objects rejected by different cuts indicated by different line colours as labelled: blue shows \GAMA objects missed due the star-galaxy separation applied (\SG), green due to large galaxy masking (\LG), yellow -- bright star masking (\BS), red --  number of observations (\NOBS) and purple due to the remaining cuts (\CC, \FMC and all the \QCs). The dashed gray line shows the redshift distribution of \BGS galaxies cross-matched with \GAMA. The vertical black dashed line marks the redshift boundary between stars ($z<0.002$) and galaxies.}
    \label{fig:rejs_gama_z}
\end{figure}

 In Fig.~\ref{fig:rejs_gama_z} we show the redshift distribution of the \GAMA galaxies that are not present in the \BGS. The solid coloured lines show the distribution for \GAMA galaxies rejected by different \BGS cuts, as labelled in the figure. We also plot the overall redshift distribution of \BGS galaxies for comparison. \GAMA galaxies removed by the bright star masking and by the restrictions on the number of observations have a similar redshift distribution to the overall \BGS. \GAMA galaxies that are removed by the large galaxy mask have a distribution that is shifted to lower redshifts than the overall \BGS distribution. \GAMA galaxies can be found within the geometric \BGS mask as \GAMA does not use masking to deal with large galaxies, and so \GAMA galaxies can be found in the regions that the \BGS rejects around large galaxies. However, \GAMA does perform masking around  bright stars but this is less aggressive than the \LS \DReight bright star masking. This can be seen from the areas rejected: the bright star masking in \GAMA removes $\sim1$ object/deg$^2$ \citep{10.1111/j.1365-2966.2010.16282.x} whereas LS \DReight removes $\sim5$ objects/deg$^2$.

\subsection{Potential systematics}\label{subsec:systematics}

Here we look at potential systematic effects that could influence the homogeneity of the \BGS catalogue and show how to mitigate these. 
As in any survey, the density of BGS targets is affected by observational effects which arise for a number of reasons.
These include astrophysical foregrounds such as Galactic extinction, variations in the density of stars in the Milky Way, as well as variations in depth for the different imaging surveys and uncertainties in the data calibration.

\begin{figure*}
	\includegraphics[width=18cm]{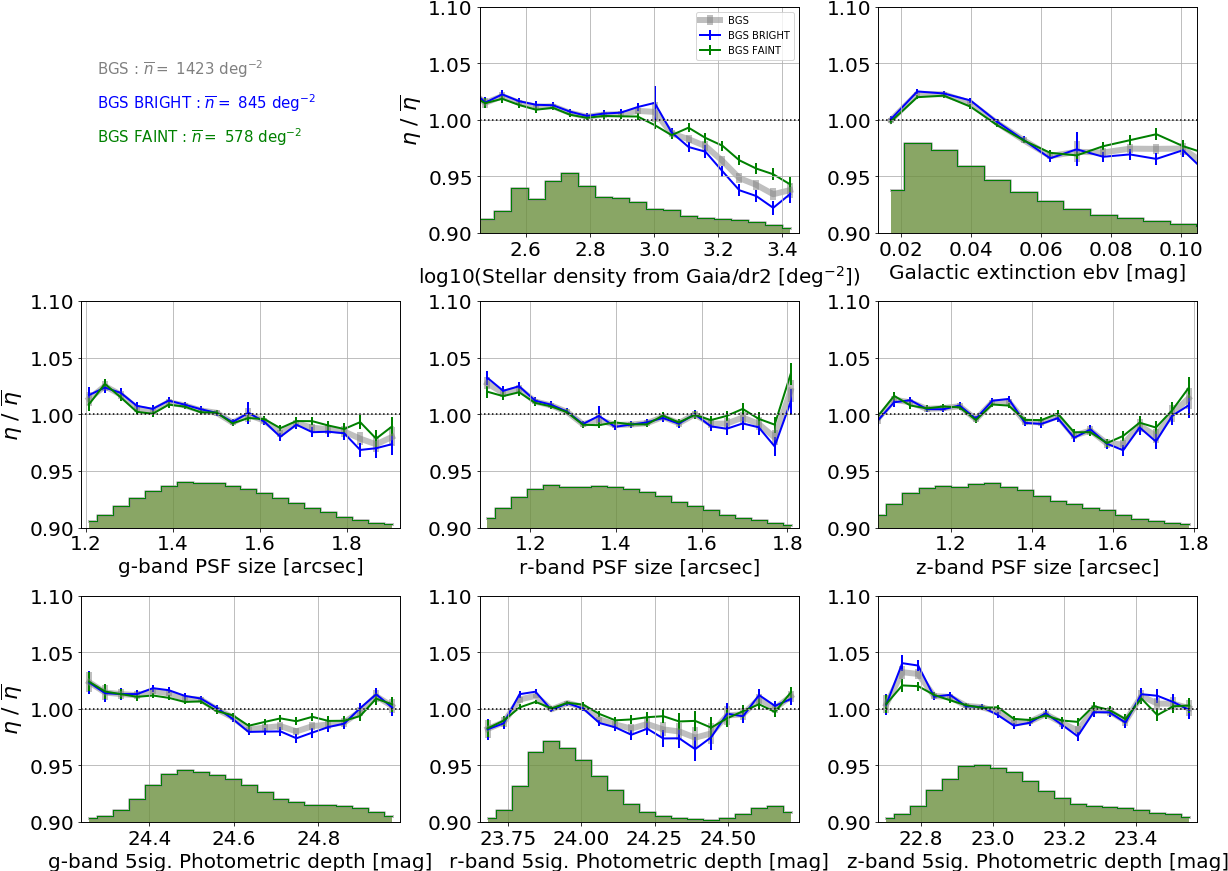}
    \caption{The systematic variation of the \BGSB (blue) and \BGSF (green) and  combined (bgs\_any, gray) target densities with respect to different properties: the logarithm of the stellar density from \GAIA DR2, Galactic extinction, \PSF size in the three bands ($grz$) and the photometric depth in each of the three bands ($grz$).
    The target densities and these eight quantities were computed in pixels on the sky using a HEALPix grid with resolution of $N_{\rm side} = 256$. Histograms shows the distribution for each of the x-axis properties. The error bars show the errors on the mean. Each target density, $\eta$ is expressed in units of its mean across the whole survey $\bar \eta$ as given in the legend.}
    \label{fig:sys_unweighted}
\end{figure*}

To study the impact of these systematics on the observed galaxy density, we use a HEALPix map that divides the whole sky into $12N_{\rm side}^2$ equal area pixels, adopting $N_{\rm side} = 256$. Each pixel contains the median value of the systematics values within the pixel and the \BGS target density. The corresponding \BGS target density in each pixel, $\eta_{\rm i}$, is defined in Equation~\ref{eq:eta_sys}.


We study the effect of eight systematics on the \BGS target density:
\begin{itemize}
\item Stellar density: 
we use stars from the \GAIA DR2 catalogue with $12<G<17$ to construct the stellar density in each HEALPix pixel.
\item Galactic extinction: the extinction  values were computed using the \textsc{sfd98} dust maps as reviewed in  Section~\ref{subsec:lsdr8}.
\item \PSF size (seeing) in the $grz$ bands: 
the \PSF size measures the full width at half maximum (FWHM) of the point spread function (PSF) which determines how much the transmission of light through turbulence in the Eath's atmosphere blurs the observed images. The seeing varies across the multiple observations.

\item Photometric depth in the $grz$ bands: the depth of the photometry, as characterised by the 5$\sigma$ AB magnitude detection limit for a $0.45$ arcsec round exponential galaxy profile, varies across the survey due to changes in the observing conditions. 
\end{itemize}

To determine if the \BGS target density has a systematic dependence on any of these quantities, we bin the HEALPix pixels according to the value of the quantity and for each bin determine the mean target density, $\eta_{i}$, and the error on the mean, $\sigma_{i}/\sqrt{N_{i}}$. In Fig.~\ref{fig:sys_unweighted} we show how the mean \BGS target density, $\eta$, varies with respect to each of the quantities listed above. Each panel shows the mean and error on the mean for three samples, \BGSB, \BGSF  and the combined \BGS sample (labelled bgs\_any). The histogram below the curves in each panel shows (on an arbitrary scale) the number of HEALPix pixels contributing to each estimate. In general, the systematic variation in the \BGS target density is less than $5$ per cent, with the one exception being a $\sim7$ per cent decrease in the target density in regions of high stellar density. 
  
Stars could impact the \BGS target density in at least five ways:
 i) Stellar contamination of the \BGS selection could lead to increased target density in regions of the sky with high stellar density. ii) While the impact of very bright stars is dealt with by masking (see Section~\ref{subsubsec:BS}), the halos and diffraction spikes around slightly fainter stars could still affect the photometry of neighbouring galaxies. iii) High stellar density could lead to an overestimate of the local sky brightness which, when subtracted, would lead to fainter galaxy fluxes and hence a lower \BGS target density. iv) star/galaxy superposition. v) Binary stars that \TRACTORs resolution is not capable of resolving.

Stellar contamination would lead to an increase in target density with increasing stellar density, whereas we see a decrease that sets in above a stellar density of  $10^{3}$~deg$^{-2}$. Hence, stellar contamination cannot be the dominant systematic influence on the target density.  

Galaxy photometry directly compromised by nearby stars that were not subject to masking also seems unlikely to be the cause for the variation in target density. We test this by implementing the medium bright stars mask with a very little impact on target density and clustering. A further masking with $2$ and $3$ times the masking radius of equation (\ref{eq:mag_radii_bs}) was also tested with no improvement on target density at high stellar densities.


The effect of high stellar density on the estimation of the sky levels deserves further investigation, but is deferred to another study. 
There is some variation of the target density with galactic extinction which could indicate 
systematic errors in the estimation of the amount of dust extinction. However, as there are spatial correlations between stellar density and dust extinction, these trends could be driven by the variation in stellar density and can be mitigated with several techniques such as linear and non-linear regressions and machine learning techniques such as Artificial Neural Networks \citep{2020MNRAS.495.1613R}.

Due to variations in observing conditions, the \PSF size varies across the survey. The explicit modelling of the PSF of each image by \TRACTOR should make the photometry robust to this variation. Also, our use of \GAIA to perform star-galaxy separation should also make this classification independent to variations in the seeing. This appears to be borne out by the results shown in Fig.~\ref{fig:sys_unweighted} which exhibit only very weak trends with \PSF.

In the \BGS, while the primary selection is in the $r$-band, \TRACTOR simultaneously fits objects in all 3 bands and so the model parameters are affected by data in all three bands. However, any dependence on the depth of the photometry appears very weak in all three bands. This to be expected as the photometric depth is typically 3 to 4 magnitudes deeper than the $r=20$ selection limit of the \BGS.

\subsubsection{Mitigation of systematics using linear weights based on stellar density}\label{subsubsec:linwts}

One way to mitigate the effect of the systematics in our catalogue is to apply a weight that corrects the target density. If we treat the systematic dependence of the observed target density on a particular quantity, $S$, as a simple regression problem, we can define the  observed target density, $\eta_{\rm i}^{\rm o}$, averaged over HEALpix pixels with a  particular value of $S=S_{\rm i}$, as
\begin{equation}
    \eta_{\rm i}^{\rm o} = \eta_{\rm i} \, W_{\rm i}(S_{\rm i}).
\end{equation}
Here, $\eta_{\rm i}$ is the true target density and $W_{\rm i}(S_{\rm i})$ is the weight for a given systematic attribute, $S$. As shown in Fig.~\ref{fig:sys_unweighted}, the most important target density variation is driven by stellar density. Here, we assume that the weight is a simple linear function, $W_{\rm i}(S_{\rm i}) = m S_{\rm i} + c$, where $S_{\rm i}$ is the the stellar density, as we would expect any contamination (or anti-contamination) to be proportional to the stellar density and not to the $\log_{10}(\textrm{stellar density})$. The best fitting coefficients we find when applying this model to the combined \BGSB and \BGSF sample are $c=1.03$ and $m=-3.96\times 10^{-5}$. By construction, this weighting removes the general trend with stellar density for the combined sample and most of the trend with stellar density for the individual \BGSB and \BGSF samples.  At the same time this weighting also reduces the weak systematic trend of target density with galactic extinction.



\subsection{Angular correlation function}\label{subsec:angular_correlation}

We measure the angular correlation function, $w(\theta)$, in five apparent magnitude bins from $r_{\rm AB}=15$ to $r_{\rm AB}=20$ for the \BGS targets in \DECaLS South Galactic Cap (\SGC) and North Galactic Cap (\NGC). Angular correlations were computed using the publicly available code CUTE \citep{2012arXiv1210.1833A}. We compare these with measurements from the mock \BGS lightcone catalogue~\citep{Smith:2017tzz}. This mock catalogue was built by populating the MXXL N-body simulation with galaxies based on a halo occupation distribution model. By construction, the HOD parameters of this mock reproduces both the luminosity function and 2-point clustering measured in the \SDSS at low redshift and the GAMA survey at higher redshift.

\begin{figure}
    \centering
	\includegraphics[width=\columnwidth]{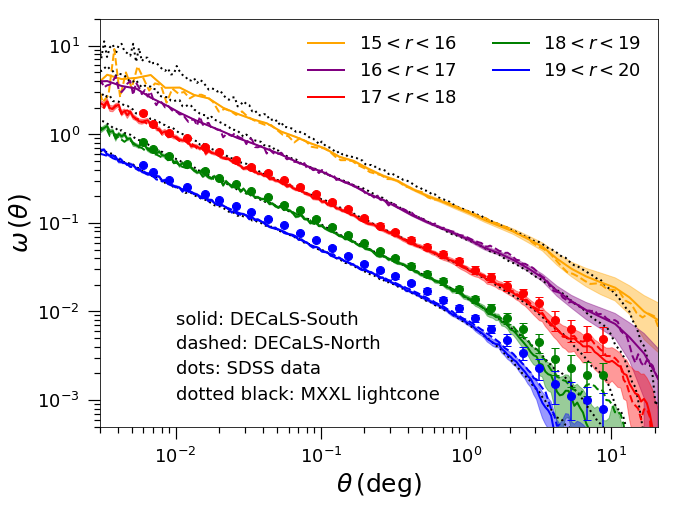}
    \caption{The angular correlation function, $w(\theta)$, measured for the \BGS targets in bins of apparent magnitude; different colours indicate different magnitude bins as labelled. The shaded area shows the standard deviation obtained from 100 jackknife regions. The solid curves show the results for DECaLS-South, the dashed curves show  DECaLS-North and the dotted curves show the angular clustering in the MXXL lightcone catalogue. The symbols with error bars show measurements from the \SDSS by \protect\cite{Wang:2013noa}.}
    \label{fig:angular-noweight}
\end{figure}

Fig.~\ref{fig:angular-noweight} shows the comparison of angular clustering measured for the BGS targets with error bars corresponding to the standard deviation of 100 jackknife realisations, the MXXL mock and the \SDSS observations by \cite{Wang:2013noa}. 
The angular clustering measurements are consistent between the DECaLS North and South regions, which demonstrates the homogeneity between these two parts of DECaLS. The angular clustering of the BGS targets agrees very well with that displayed in the MXXL lightcone. The HOD parameters of the MXXL mock have been tuned to attempt to match the clustering measured from \SDSS MGS, however on large scales HOD models can only alter the amplitude and not  the shape of the correlation. Moreover the shape of the large scale correlation function of MXXL is very similar to that of all \LCDM models that are consistent with CMB observations. Hence it is interesting that for the two faintest bins BGS is more consistent with MXXL (and hence with \LCDM) than is \SDSS MGS -- possibly indicating reduced systematic errors.

We also look at the angular clustering of the BGS targets after applying the weights that depend on stellar density, as described in the previous section. Overall, applying stellar density weights has a small impact at angular scales larger than $3-4$ deg. Both the clustering with and without the weights are consistent with each other, within the error bar.

A further test of the fidelity of our \BGS catalogue is to check for any spatial correlation of the distribution of \BGS targets with stars in the Milky Way. Here we focus our attention on the fainter stars, 12$<G<$17, which, ideally, should be removed from the \BGS targets by our star-galaxy 
separation scheme. We find a significant anticorrelation on very small scales 
but no correlation on scales larger than 100 arc seconds.

\subsection{Angular cross-correlation with large galaxies}\label{subsec:angular_cross_lg}

\begin{figure}
    \centering
	\includegraphics[width=\columnwidth]{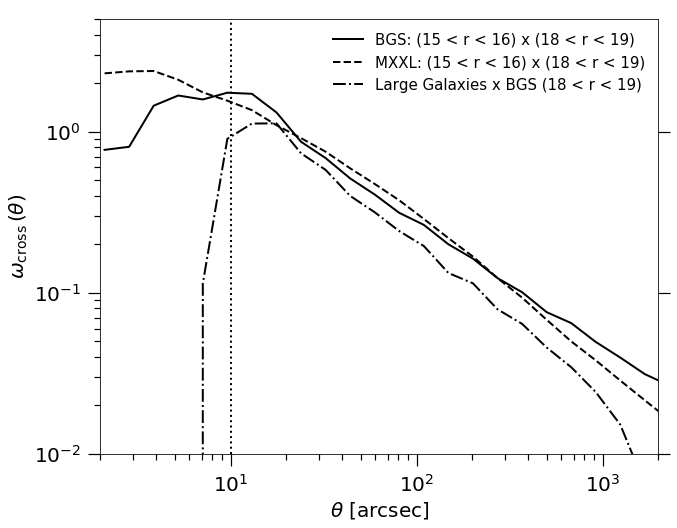}
    \caption{The angular cross-correlation function measured between faint BGS targets in $18 < r < 19$ and  large galaxies from the \SGA-2020 (dashdotted) and between the same faint BGS targets and brighter BGS targets in $15< r < 16$ (solid), the magnitude range in which most of the large galaxies reside. We also compare with the angular cross-correlation between these two bins in apparent magnitude measured in the MXXL lightcone (dashed). The vertical dotted line shows the mean LG mask radius which is about 10 arcsec.}
    \label{fig:xlg}
\end{figure}

In order to determine whether we are missing faint \BGS targets around large galaxies due to the LG mask defined in Section~\ref{subsec:SGA}, we measure the angular cross-correlation function between the \SGA-2020 and faint \BGS targets in $18 < r < 19$ (dash-dotted) as shown in Fig.~\ref{fig:xlg}. We also measure the angular cross-correlation function between these faint \BGS targets and brighter \BGS targets in the magnitude range $15< r < 16$ (solid) where we assume that most of the large galaxies lie, and we do the same using the MXXL lightcone (dashed). The vertical dotted line shows the mean mask radius around large galaxies, which is about 10~arcsec.

The agreement between the results from the \BGS catalogue (solid) and from the MXXL lightcone (dashed) suggests that our treatment of large galaxies is satisfactory and we are only missing \BGS targets on scales below 10 arcsec,
which is the median large galaxy masking radius (see Section~\ref{subsubsec:LG}).
The difference in amplitude between the solid and dash-dotted curves, with a lower value when cross-correlating with the \SGA-2020, suggests that the catalogue of large galaxies contains either more low-z galaxies or more brighter galaxies, or both, compared to the \BGS targets in $15< r < 16$.   

\section{Conclusions}\label{sec:conclusions}

Here we have presented the steps needed to define and select the Bright Galaxy Survey (\BGS) targets for the Dark Energy Spectroscopic Instrument (DESI) project. Our galaxy selection uses \DECaLS \LS imaging data from Data Release 8 (\DReight) reduced by the \NOAO \CP and \TRACTOR pipelines. Our \BGS target selection has two main components, one which imposes spatial cuts and the other which applies photometric selections. Figs.~\ref{fig:flow1} and~\ref{fig:flow2} show the flowcharts that set out these two selections. At each step these flowcharts report the remaining survey area and surface density of targets.   
 
The main features of our spatial and photometric cuts are the following:

\begin{itemize}

\item The \BGS spatial target selection removes area near bright stars (\BS mask), large galaxies (\LG mask), and globular clusters (\GC mask), as well as galaxies with less than a specified minimum number of observations (\NOBS mask). The \BS mask is a circular aperture that scales with the magnitude of the bright star (see Eqn.~\ref{eq:mag_radii_bs}). The exclusion of areas around bright stars removes $\sim270$ deg$^2$, this is $2.76$ per cent of initial footprint. Inspection of stacked images around bright stars (i.e. those with \GAIA \, $G < 13$ or \TYCHO \, $V < 13$) shows that the \BS masking radius used in \TRACTOR is well-motivated, with no sign of contamination around the bright stars in the \BGS target density. There is a modest $\sim6$ per cent increase in BGS target density just beyond the edge of the masked region. We find that there is a negligible angular cross-correlation between stars and galaxies at scales $> 100$ arcsec. Below $100$ arcsec we have an anti-correlation possibly caused by the stars masked within the range $12 < G < 13$.

\item The \LG and \GC masks account for a smaller number of contaminants than the \BS mask, removing just $\sim9$ deg$^2$ of survey area or $0.09$ percent of initial footprint. 

\item \DECaLS \DReight is complete to $99.5$ per cent with at least one observation in the three bands $grz$, as described by the value of \NOBS. The selection made on \NOBS removes $\sim39$ deg$^2$ of imaging data.


\item We use \GAIA DR2 to separate stars and galaxies as described in Section~\ref{subsec:star_galaxy}. This classification exploits the small PSF of the \GAIA imaging compared with that typically present in ground-based observations. In our classification scheme we compare the measurement of the flux of an object by \GAIA with that from \TRACTOR through the parameter $G-rr$. Objects with a \TRACTOR flux that is greater than that reported by \GAIA are considered to be galaxies because this difference implies that they are  extended sources (see Fig.~\ref{fig:Grr-gz}).

\item A small fraction ($\sim0.35$ per cent) of \BGS galaxies are of PSF type according to \TRACTOR. About half of these are compact sources for which the PSF model is the best fit, but the other half have only PSF photometry as they  were designated stars based on the \GAIA Astrometric Excess Noise (AEN) parameter before \TRACTOR was run. For these objects \TRACTOR only performs PSF fits.  Matching to \GAMA reveals that most (96 per cent) of these \BGS PSF-type objects are confirmed to be galaxies by the \GAMA spectroscopy. In addition, we find that the 
$\sim7$ \GAMA galaxies/deg$^2$ that are missed in \BGS are mostly ($\sim98$ per cent) of PSF type according to \TRACTOR .  We conclude that using the AEN classification is i) causing $\sim0.17$ per cent of \BGS galaxies to be falsely classfied as of PSF type and ii) compromising the photometry of another $7$ objects/deg$^2$ which then due to having their fluxes underestimated are falsely classified as stars by the \BGS  $G-rr$ star-galaxy classification.

\item Possible systematic effects in \DECaLS leave a small imprint on surface density of \BGS sources. The variation in the target density of \BGS sources as a function of the main possible systematic effects, such as the stellar density, galactic extinction, seeing and imaging depth, is less than 10 per cent in the case of stellar density and under 5 per cent for the remaining systematics. We implement a weighting scheme based on a linear regression model which uses the density of stars to mitigate these effects. Applying the resulting weights, variation in the target density with stellar density is removed by construction, and is greatly reduced when plotted against the other systematic quantities. 

\item Angular clustering measurements made from our \BGS target catalogue are compared with previous measurements from \SDSS and the predictions from the \MXXL lightcone mock catalogue, which on large scales can be taken as a prediction of \LCDM models (see~\ref{subsec:angular_correlation}). On small scales, the three measurements of the angular correlation function agree well, with the exception of the brightest galaxies considered. At large scales, the angular clustering we find for the \BGS targets is closer to that recovered from the \MXXL mock catalogue than the \SDSS measurements. The agreement between the \BGS and the MXXL lightcone is even better after applying the linear weights based on stellar density to the \BGS.

\end{itemize}


Galleries with examples of \BGS targets divided in \BGSB and \BGSF can be found at \url{http://astro.dur.ac.uk/\~qmxp55/bgs_ts_paper_gallery.html} along with galleries showing examples of rejected objects by the different spatial and photometric cuts we apply in \BGS. We included also examples of discrepancies between our star-galaxy (\SG) classification using \GAIA with \TRACTORs divided into 1) \TRACTORs extended objects that fail our \SG classification, and the \TRACTORs point sources objects that pass our \SG classification and 2) are \GAIA and 3) are not \GAIA sources. Finally, examples of discrepancies between \TRACTORs point source classification for \GAIA objects and our \SG classification divided in two samples: 1) are galaxies by our SG classification but stars according to \TRACTORs assessment of \GAIA sources using the Astrometric Excess Noise (AEN) parameter from \GAIA, and 2) stars by our \SG classification but galaxies by their AEN classification.

In a second paper we will focus on applying this framework to select \BGS targets using the additional  \LS, \BASS and \MzLS imaging data, and set out what is needed to tune our selection to use the upcoming release of the \LS, DR9. Among the main changes in DR9 compared to DR8 are 
i) the implementation of an iterative source detection process in \TRACTOR
in which the detection algorithm is rerun after sources have been fitted and subtracted,
ii) an extended \PSF model to subtract the wings of bright stars, iii) the COMPOSITE (\COMP) \TRACTOR model has been replaced with a SERSIC model, (\SER) iv) the criteria used to determine which \GAIA objects are forced to be fitted by the \PSF model 
are now more restrictive, v) adjustments have been made to the masking procedure around bright stars and to fainter MEDIUM stars where the masking radius around bright stars has been reduced by a factor of two. In addition, TRACTOR implements a local fit to the sky background around these objects. vi) \SGA-2020 and Globular Cluster catalogues have been updated and the large galaxy photometry redone in their own custom run of \TRACTOR. It is expected that (i) will marginally increase the completeness of the \BGS catalogue,  (iv) will reduce the incidence of galaxies being misclassified as stars, and the other changes will improve the photometry. 
A second paper will quantify these changes and focus 
predominately on changes in selection relative to this DR8 selection. Hence most of the details of the \BGS selection will be in this paper only. Despite these improvements in the quality of the selection, early test releases of the upcoming DR9 data suggest that \BGS targets will not vary more than $5$ per cent compared to present selection with \DECaLS DR8. The second paper will also include a more complete clustering analysis using mock catalogues and colour based clustering measurements, and a more sophisticated technique for the mitigation of systematic effects. A third paper we will cover the work we have undertaken to define and select the \BGS targets for the survey validation programme.This series of papers is intended to be complementary work to the overall \DESI key project paper on target selection aimed to be released in 2021.

\section*{Acknowledgements}
We acknowledge helpful conversations with Anand Raichoor and Christophe Yeche. OR-M is supported by the Mexican National Council of Science and Technology (CONACyT) through grant No. 297228/440775 and funding from the European Union’s Horizon 2020 Research and Innovation Programme under the Marie Sklodowska-Curie grant agreement No 734374. SC, PN, PZ, CMB and JL acknowledge support from the Science Technology Facilities Council through ST/P000541/1 and ST/T000244/1. ADM was supported by the U.S. Department of Energy, Office of Science, Office of High Energy Physics, under Award Number DE-SC0019022. JM gratefully acknowledges support from the U.S. Department of Energy, Office of Science, Office of High Energy Physics under Award Number DE-SC002008 and from the National Science Foundation under grant AST-1616414.

This research used resources of the National Energy Research Scientific Computing Center (NERSC), a U.S. Department of Energy Office of Science User Facility operated under Contract No. DEAC02-05CH11231. This work also made extensive use of the NASA Astrophysics Data System and of the astro-ph preprint archive at arXiv.org. Authors want to thank the GAMA collaboration for early access to GAMA DR4 data for this work. Some of the results in this paper have been derived using the healpy and HEALPix package. We acknowledge the usage of the HyperLeda database (\url{http://leda.univ-lyon1.fr}).

This work used the DiRAC@Durham facility managed by the Institute for Computational Cosmology on behalf of the STFC DiRAC HPC Facility (www.dirac.ac.uk). The equipment was funded by BEIS capital funding via STFC capital grants ST/K00042X/1, ST/P002293/1 and ST/R002371/1, Durham University and STFC operations grant ST/R000832/1. DiRAC is part of the National e-Infrastructure. 

This research is supported by the Director, Office of Science, Office of High Energy Physics of the U.S. Department of Energy under Contract No. DE–AC02–05CH1123, and by the National Energy Research Scientific Computing Center, a DOE Office of Science User Facility under the same   contract; additional support for DESI is provided by the U.S. National Science Foundation, Division of Astronomical Sciences under Contract No. AST-0950945 to the NSF’s National Optical-Infrared Astronomy Research Laboratory; the Science and Technologies Facilities Council of the United Kingdom; the Gordon and Betty Moore Foundation; the Heising-Simons Foundation; the French Alternative Energies and Atomic Energy Commission (CEA); the National Council of Science and Technology of Mexico; the Ministry of Economy of Spain, and by the DESI Member Institutions.  The authors are honored to be permitted to conduct astronomical research on Iolkam Du\textquotesingle ag (Kitt Peak), a mountain with particular significance to the Tohono O \textquotesingle odham Nation.

\section*{Data availability}

The data used in this paper is publicly available. The preliminary \BGS target selection described in this paper is public at \url{https://data.desi.lbl.gov/public/ets/target/catalogs/} and detailed at \url{https://desidatamodel.readthedocs.io}.
The DESI Legacy Imaging Surveys used for this work is public at \url{https://www.legacysurvey.org/dr8/description/}.




\bibliographystyle{mnras}
\bibliography{sample} 

\begin{thebibliography}{}
\makeatletter
\relax
\def\mn@urlcharsother{\let\do\@makeother \do\$\do\&\do\#\do\^\do\_\do\%\do\~}
\def\mn@doi{\begingroup\mn@urlcharsother \@ifnextchar [ {\mn@doi@}
  {\mn@doi@[]}}
\def\mn@doi@[#1]#2{\def\@tempa{#1}\ifx\@tempa\@empty \href
  {http://dx.doi.org/#2} {doi:#2}\else \href {http://dx.doi.org/#2} {#1}\fi
  \endgroup}
\def\mn@eprint#1#2{\mn@eprint@#1:#2::\@nil}
\def\mn@eprint@arXiv#1{\href {http://arxiv.org/abs/#1} {{\tt arXiv:#1}}}
\def\mn@eprint@dblp#1{\href {http://dblp.uni-trier.de/rec/bibtex/#1.xml}
  {dblp:#1}}
\def\mn@eprint@#1:#2:#3:#4\@nil{\def\@tempa {#1}\def\@tempb {#2}\def\@tempc
  {#3}\ifx \@tempc \@empty \let \@tempc \@tempb \let \@tempb \@tempa \fi \ifx
  \@tempb \@empty \def\@tempb {arXiv}\fi \@ifundefined
  {mn@eprint@\@tempb}{\@tempb:\@tempc}{\expandafter \expandafter \csname
  mn@eprint@\@tempb\endcsname \expandafter{\@tempc}}}

\bibitem[\protect\citeauthoryear{Abazajian et~al.}{Abazajian
  et~al.}{2003}]{Abazajian:2003jy}
Abazajian K.,  et~al., 2003, \mn@doi [Astron. J.] {10.1086/378165}, 126, 2081

\bibitem[\protect\citeauthoryear{{Adelman-McCarthy} et~al.,}{{Adelman-McCarthy}
  et~al.}{2008}]{2008ApJS..175..297A}
{Adelman-McCarthy} J.~K.,  et~al., 2008, \mn@doi [\apjs] {10.1086/524984},
  \href {https://ui.adsabs.harvard.edu/abs/2008ApJS..175..297A} {175, 297}

\bibitem[\protect\citeauthoryear{{Alonso}}{{Alonso}}{2012}]{2012arXiv1210.1833A}
{Alonso} D.,  2012, arXiv e-prints, \href
  {https://ui.adsabs.harvard.edu/abs/2012arXiv1210.1833A} {p. arXiv:1210.1833}

\bibitem[\protect\citeauthoryear{{Angulo}, {Springel}, {White}, {Jenkins},
  {Baugh}  \& {Frenk}}{{Angulo} et~al.}{2012}]{Angulo2012}
{Angulo} R.~E.,  {Springel} V.,  {White} S.~D.~M.,  {Jenkins} A.,  {Baugh}
  C.~M.,   {Frenk} C.~S.,  2012, \mn@doi [\mnras]
  {10.1111/j.1365-2966.2012.21830.x}, \href
  {https://ui.adsabs.harvard.edu/abs/2012MNRAS.426.2046A} {426, 2046}

\bibitem[\protect\citeauthoryear{{Bailer-Jones}, {Fouesneau}  \&
  {Andrae}}{{Bailer-Jones} et~al.}{2019}]{2019MNRAS.490.5615B}
{Bailer-Jones} C. A.~L.,  {Fouesneau} M.,   {Andrae} R.,  2019, \mn@doi
  [\mnras] {10.1093/mnras/stz2947}, \href
  {https://ui.adsabs.harvard.edu/abs/2019MNRAS.490.5615B} {490, 5615}

\bibitem[\protect\citeauthoryear{Baldry et~al.,}{Baldry
  et~al.}{2010}]{10.1111/j.1365-2966.2010.16282.x}
Baldry I.~K.,  et~al., 2010, \mn@doi [Monthly Notices of the Royal Astronomical
  Society] {10.1111/j.1365-2966.2010.16282.x}, 404, 86

\bibitem[\protect\citeauthoryear{Baldry et~al.,}{Baldry
  et~al.}{2017}]{10.1093/mnras/stx3042}
Baldry I.~K.,  et~al., 2017, \mn@doi [Monthly Notices of the Royal Astronomical
  Society] {10.1093/mnras/stx3042}, 474, 3875

\bibitem[\protect\citeauthoryear{{Bertin} \& {Arnouts}}{{Bertin} \&
  {Arnouts}}{1996}]{1996A&AS..117..393B}
{Bertin} E.,  {Arnouts} S.,  1996, \mn@doi [\aaps] {10.1051/aas:1996164}, \href
  {https://ui.adsabs.harvard.edu/abs/1996A&AS..117..393B} {117, 393}

\bibitem[\protect\citeauthoryear{{Blanton} et~al.,}{{Blanton}
  et~al.}{2001}]{2001AJ....121.2358B}
{Blanton} M.~R.,  et~al., 2001, \mn@doi [\aj] {10.1086/320405}, \href
  {https://ui.adsabs.harvard.edu/abs/2001AJ....121.2358B} {121, 2358}

\bibitem[\protect\citeauthoryear{{Carrasco} et~al.,}{{Carrasco}
  et~al.}{2016}]{2016A&A...595A...7C}
{Carrasco} J.~M.,  et~al., 2016, \mn@doi [\aap] {10.1051/0004-6361/201629235},
  \href {https://ui.adsabs.harvard.edu/abs/2016A&A...595A...7C} {595, A7}

\bibitem[\protect\citeauthoryear{{Chambers} et~al.,}{{Chambers}
  et~al.}{2016}]{2016arXiv161205560C}
{Chambers} K.~C.,  et~al., 2016, arXiv e-prints, \href
  {https://ui.adsabs.harvard.edu/abs/2016arXiv161205560C} {p. arXiv:1612.05560}

\bibitem[\protect\citeauthoryear{{DESI Collaboration} et~al.,}{{DESI
  Collaboration} et~al.}{2016}]{DESI2016:surveys}
{DESI Collaboration} et~al., 2016, arXiv e-prints, \href
  {https://ui.adsabs.harvard.edu/abs/2016arXiv161100036D} {p. arXiv:1611.00036}

\bibitem[\protect\citeauthoryear{Dey et~al.,}{Dey et~al.}{2019}]{Dey:2019}
Dey A.,  et~al., 2019, \mn@doi [The Astronomical Journal]
  {10.3847/1538-3881/ab089d}, 157, 168

\bibitem[\protect\citeauthoryear{{Driver} et~al.,}{{Driver}
  et~al.}{2012}]{2012yCat..74130971D}
{Driver} S.~P.,  et~al., 2012, VizieR Online Data Catalog, \href
  {https://ui.adsabs.harvard.edu/abs/2012yCat..74130971D} {p. J/MNRAS/413/971}

\bibitem[\protect\citeauthoryear{{Fadely}, {Hogg}  \& {Willman}}{{Fadely}
  et~al.}{2012}]{2012ApJ...760...15F}
{Fadely} R.,  {Hogg} D.~W.,   {Willman} B.,  2012, \mn@doi [\apj]
  {10.1088/0004-637X/760/1/15}, \href
  {https://ui.adsabs.harvard.edu/abs/2012ApJ...760...15F} {760, 15}

\bibitem[\protect\citeauthoryear{{Flaugher} et~al.,}{{Flaugher}
  et~al.}{2015}]{2015AJ....150..150F}
{Flaugher} B.,  et~al., 2015, \mn@doi [\aj] {10.1088/0004-6256/150/5/150},
  \href {https://ui.adsabs.harvard.edu/abs/2015AJ....150..150F} {150, 150}

\bibitem[\protect\citeauthoryear{{Gaia Collaboration} et~al.,}{{Gaia
  Collaboration} et~al.}{2016a}]{2016A&A...595A...1G}
{Gaia Collaboration} et~al., 2016a, \mn@doi [Astronomy and Astrophysics]
  {10.1051/0004-6361/201629272}, \href
  {https://ui.adsabs.harvard.edu/abs/2016A&A...595A...1G} {595, A1}

\bibitem[\protect\citeauthoryear{{Gaia Collaboration} et~al.,}{{Gaia
  Collaboration} et~al.}{2016b}]{2016A&A...595A...2G}
{Gaia Collaboration} et~al., 2016b, \mn@doi [Astronomy and Astrophysics]
  {10.1051/0004-6361/201629512}, \href
  {https://ui.adsabs.harvard.edu/abs/2016A&A...595A...2G} {595, A2}

\bibitem[\protect\citeauthoryear{{Gaia Collaboration} et~al.,}{{Gaia
  Collaboration} et~al.}{2018}]{2018A&A...616A...1G}
{Gaia Collaboration} et~al., 2018, \mn@doi [Astronomy and Astrophysics]
  {10.1051/0004-6361/201833051}, \href
  {https://ui.adsabs.harvard.edu/abs/2018A&A...616A...1G} {616, A1}

\bibitem[\protect\citeauthoryear{{H{\o}g} et~al.,}{{H{\o}g}
  et~al.}{2000}]{2000A&A...355L..27H}
{H{\o}g} E.,  et~al., 2000, \aap, \href
  {http://adsabs.harvard.edu/abs/2000A%26A...355L..27H} {355, L27}

\bibitem[\protect\citeauthoryear{Kitanidis et~al.,}{Kitanidis
  et~al.}{2020}]{10.1093/mnras/staa1621}
Kitanidis E.,  et~al., 2020, \mn@doi [Monthly Notices of the Royal Astronomical
  Society] {10.1093/mnras/staa1621}, 496, 2262

\bibitem[\protect\citeauthoryear{{Lang}, {Hogg}  \& {Mykytyn}}{{Lang}
  et~al.}{2016}]{2016ascl.soft04008L}
{Lang} D.,  {Hogg} D.~W.,   {Mykytyn} D.,  2016, {The Tractor: Probabilistic
  astronomical source detection and measurement}, Astrophysics Source Code
  Library (\mn@eprint {ascl} {1604.008})

\bibitem[\protect\citeauthoryear{{Liske} et~al.,}{{Liske}
  et~al.}{2015}]{2015MNRAS.452.2087L}
{Liske} J.,  et~al., 2015, \mn@doi [\mnras] {10.1093/mnras/stv1436}, \href
  {http://adsabs.harvard.edu/abs/2015MNRAS.452.2087L} {452, 2087}

\bibitem[\protect\citeauthoryear{{Makarov}, {Prugniel}, {Terekhova}, {Courtois}
   \& {Vauglin}}{{Makarov} et~al.}{2014}]{2014A&A...570A..13M}
{Makarov} D.,  {Prugniel} P.,  {Terekhova} N.,  {Courtois} H.,   {Vauglin} I.,
  2014, \mn@doi [\aap] {10.1051/0004-6361/201423496}, \href
  {http://adsabs.harvard.edu/abs/2014A%26A...570A..13M} {570, A13}

\bibitem[\protect\citeauthoryear{{Odewahn}, {Stockwell}, {Pennington},
  {Humphreys}  \& {Zumach}}{{Odewahn} et~al.}{1992}]{1992AJ....103..318O}
{Odewahn} S.~C.,  {Stockwell} E.~B.,  {Pennington} R.~L.,  {Humphreys} R.~M.,
  {Zumach} W.~A.,  1992, \mn@doi [\aj] {10.1086/116063}, \href
  {https://ui.adsabs.harvard.edu/abs/1992AJ....103..318O} {103, 318}

\bibitem[\protect\citeauthoryear{{Rezaie}, {Seo}, {Ross}  \&
  {Bunescu}}{{Rezaie} et~al.}{2020}]{2020MNRAS.495.1613R}
{Rezaie} M.,  {Seo} H.-J.,  {Ross} A.~J.,   {Bunescu} R.~C.,  2020, \mn@doi
  [\mnras] {10.1093/mnras/staa1231}, \href
  {https://ui.adsabs.harvard.edu/abs/2020MNRAS.495.1613R} {495, 1613}

\bibitem[\protect\citeauthoryear{Ruiz-Macias et~al.,}{Ruiz-Macias
  et~al.}{2020}]{Ruiz_Macias_2020}
Ruiz-Macias O.,  et~al., 2020, \mn@doi [Research Notes of the {AAS}]
  {10.3847/2515-5172/abc25a}, 4, 187

\bibitem[\protect\citeauthoryear{{Schlafly} \& {Finkbeiner}}{{Schlafly} \&
  {Finkbeiner}}{2011}]{2011ApJ...737..103S}
{Schlafly} E.~F.,  {Finkbeiner} D.~P.,  2011, \mn@doi [\apj]
  {10.1088/0004-637X/737/2/103}, \href
  {http://adsabs.harvard.edu/abs/2011ApJ...737..103S} {737, 103}

\bibitem[\protect\citeauthoryear{{Schlegel}, {Finkbeiner}  \&
  {Davis}}{{Schlegel} et~al.}{1998}]{1998ApJ...500..525S}
{Schlegel} D.~J.,  {Finkbeiner} D.~P.,   {Davis} M.,  1998, \mn@doi [\apj]
  {10.1086/305772}, \href {http://adsabs.harvard.edu/abs/1998ApJ...500..525S}
  {500, 525}

\bibitem[\protect\citeauthoryear{{Secrest}, {Dudik}, {Dorland}, {Zacharias},
  {Makarov}, {Fey}, {Frouard}  \& {Finch}}{{Secrest}
  et~al.}{2015}]{2015ApJS..221...12S}
{Secrest} N.~J.,  {Dudik} R.~P.,  {Dorland} B.~N.,  {Zacharias} N.,  {Makarov}
  V.,  {Fey} A.,  {Frouard} J.,   {Finch} C.,  2015, \mn@doi [\apjs]
  {10.1088/0067-0049/221/1/12}, \href
  {https://ui.adsabs.harvard.edu/abs/2015ApJS..221...12S} {221, 12}

\bibitem[\protect\citeauthoryear{Smith, Cole, Baugh, Zheng, Angulo, Norberg  \&
  Zehavi}{Smith et~al.}{2017}]{Smith:2017tzz}
Smith A.,  Cole S.,  Baugh C.,  Zheng Z.,  Angulo R.,  Norberg P.,   Zehavi I.,
   2017, \mn@doi [Mon. Not. Roy. Astron. Soc.] {10.1093/mnras/stx1432}, 470,
  4646

\bibitem[\protect\citeauthoryear{{Strauss} et~al.,}{{Strauss}
  et~al.}{2002}]{2002AJ....124.1810S}
{Strauss} M.~A.,  et~al., 2002, \mn@doi [\aj] {10.1086/342343}, \href
  {http://adsabs.harvard.edu/abs/2002AJ....124.1810S} {124, 1810}

\bibitem[\protect\citeauthoryear{{The Dark Energy Survey Collaboration}}{{The
  Dark Energy Survey Collaboration}}{2005}]{DES:2005}
{The Dark Energy Survey Collaboration} 2005, arXiv e-prints, \href
  {https://ui.adsabs.harvard.edu/abs/2005astro.ph.10346T} {pp
  astro--ph/0510346}

\bibitem[\protect\citeauthoryear{Wang, Brunner  \& Dolence}{Wang
  et~al.}{2013}]{Wang:2013noa}
Wang Y.,  Brunner R.~J.,   Dolence J.~C.,  2013, \mn@doi [Mon. Not. Roy.
  Astron. Soc.] {10.1093/mnras/stt450}, 432, 1961

\bibitem[\protect\citeauthoryear{{Weir}, {Fayyad}  \& {Djorgovski}}{{Weir}
  et~al.}{1995}]{1995AJ....109.2401W}
{Weir} N.,  {Fayyad} U.~M.,   {Djorgovski} S.,  1995, \mn@doi [\aj]
  {10.1086/117459}, \href
  {https://ui.adsabs.harvard.edu/abs/1995AJ....109.2401W} {109, 2401}

\bibitem[\protect\citeauthoryear{Zonca, Singer, Lenz, Reinecke, Rosset, Hivon
  \& Gorski}{Zonca et~al.}{2019}]{Zonca2019}
Zonca A.,  Singer L.,  Lenz D.,  Reinecke M.,  Rosset C.,  Hivon E.,   Gorski
  K.,  2019, \mn@doi [Journal of Open Source Software] {10.21105/joss.01298},
  4, 1298

\makeatother
\end{thebibliography}



\appendix
\section{Galaxy view}\label{app:galview}

\begin{figure*}
	\includegraphics[width=10.7cm]{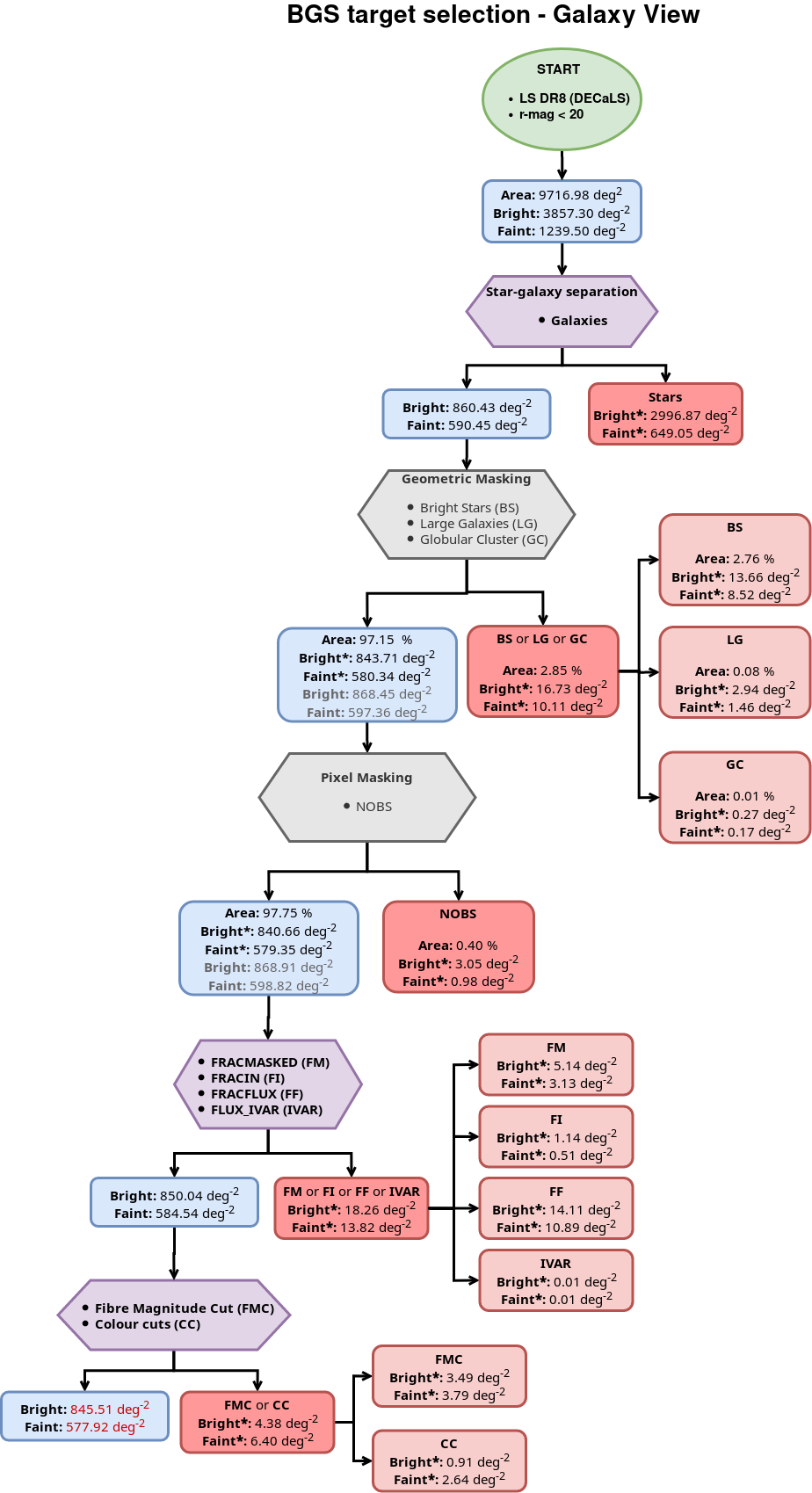}
    \caption{Flow chart showing the spatial and photometric \BGS target selections applied to the Legacy Surveys \DReight. The spatial selections are shown by gray boxes and are divided into two kinds, one defined by geometric cuts around bright sources i.e. bright stars (\BS), large galaxies (\LG) and globular clusters (\GC), and the other which is at the pixel level, such as the number of observations (\NOBS). The photometric selection of \BGS targets is divided into four types and is shown by purple boxes; star-galaxy separation, fibre magnitude cuts (\FMC), colour cuts (\CC) and quality cuts (\QCs) which include  \FRACMASKED, \FRACIN, \FRACFLUX and \FLUXIVAR. The blue boxes show the area (in degrees) and the number density (per square degree) of objects retained after each selection, broken down into the numbers for the bright and faint components of the \BGS. The red boxes show the equivalent information for the rejected objects. If more than one cut or selection is applied at a given stage, then the darker red boxes show the information about removed objects for the combination of cuts and the lighter red boxes show the corresponding values for each individual cut. The superscript ($^*$) denotes target densities without correcting for the area removed by cuts up to that point, while densities without a superscript ($^*$) do take into account the reduction in area.}
    \label{fig:flow_galaxy}
\end{figure*}

In contrast to the approach taken in the main paper, here we present a `galaxy' view of the \BGS selection by implementing the star-galaxy separation before the other \BGS cuts (with the exception of first applying the nominal \BGS magnitude limit $r< 20$). The results of this exercise are shown in Fig.~\ref{fig:flow_galaxy}. In this view, the geometric masking does not look as aggressive as it did in Fig.~\ref{fig:flow1}, with the size of the rejected area and number of objects typically reduced at each step by an order of magnitude compared to what was seen in Fig.~\ref{fig:flow1}. The \BS mask step is the stage that is the most affected by this change in order. Next is the application of the selection on \NOBS which has half the effect that it did in Fig.~\ref{fig:flow1}. Note that the area removed by the cuts remains unchanged as this does not depend on the number of targets but is calculated using the randoms.  

In addition to the changing the order in which the star-galaxy separation is applied compared to the selection criteria presented in Sections ~\ref{sec:spatial_masking} and ~\ref{sec:photo_select}, we swap the \FMC and \CC with the \QCs. When comparing both schemes,  (Fig.~\ref{fig:flow2} and Fig.~\ref{fig:flow_galaxy}), we see a high overlap between the \QCs and the \FMC of  $\sim15$  objects/deg$^{2}$ which represent $\sim2/3$ the galaxies rejected by \FMC in Section~\ref{sec:photo_select}. \CC is also affected by the to the sequence of cuts and the rejections due to this cut is reduced by a factor of $2$ in the galaxy view. 

\bsp	
\label{lastpage}
\end{document}